\documentclass{cmslatex}
\usepackage[paperwidth=7in, paperheight=10in, margin=.875in]{geometry}
 \usepackage[backref,colorlinks,linkcolor=red,anchorcolor=green,citecolor=blue]{hyperref}
\usepackage{amsfonts,amssymb}
\usepackage{amsmath}
\usepackage{amstext}
\usepackage{graphicx}
\usepackage{cite}
\usepackage{enumerate}
\usepackage[colorinlistoftodos]{todonotes}

\renewcommand{\theequation}{\arabic{section}.\arabic{equation}}

\allowdisplaybreaks

\newcommand{\rset}{\mathbb{R}}
\newcommand{\cset}{\mathbb{C}}
\newcommand{\Rd}{\mathrm{d} }
\newcommand{\bA}{\bar{A}} %
\newcommand{\TR}{\mathrm{Tr}\,} %

\newcommand{\FTERM}{\mathrm{I}}
\newcommand{\SECTERM}{\mathrm{II}}

\newcommand{\OPW}{\hat{W}}
\newcommand{\tLm}{\lambda}
\newcommand{\tVd}{V}

\def\XYHALF{\frac{1}{2}(x+y)}
\def\HZ{H}
\def\SCPRODE#1#2{\left\langle{#1},{#2}\right\rangle_{e}}
\def\LSPEL{L^2(\rset^{3J}\times\bar\Sigma)}
\def\LSP{L^2(\rset^{3(J+N)}\times\Sigma\times\bar\Sigma)}

\def\OPER#1{\hat{#1}}
\def\OPERW#1{\widehat{#1}}
\def\PERIOD{\,.}
\def\COMMA{\,,}

\def\IU{\mathrm{i}}
\def\NORMFAC{\left(\frac{\sqrt M}{2\pi}\right)}

\def\MP{{\#}}

\def\HAM{\mathcal{H}}
\def\SCPROD#1#2{\langle {#1},{#2}\rangle}

\begin{document}

\title{The classical limit of quantum observables in the conservation laws of fluid dynamics%
\thanks{Received date:}
      }

\author{Petr Plech\'{a}\v{c}%
        \thanks{Department of Mathematical Sciences, 
         University of Delaware,
         Newark, DE 19716, USA, (plechac@udel.edu)}
    \and
         Mattias Sandberg%
         \thanks{Institutionen f\"or Matematik, Kungl. Tekniska H\"ogskolan, 100 44 Stockholm, Sweden. (msandb@kth.se)}
    \and
         Anders Szepessy%
        \thanks{Institutionen f\"or Matematik, Kungl. Tekniska H\"ogskolan, 100 44 Stockholm, Sweden. (szepessy@kth.se)}
    }
\thanks{The research of A.S. and M.S. was supported by
Swedish Research Council 621-2014-4776 and the Swedish e-Science Research Center. The research of P.P. was supported by ARO MURI Award No.  W911NF-14-024}

\pagestyle{myheadings}
\markboth{Classical Limit of quantum observables in the conservation laws of fluid dynamics}{P. Plech\'a\v{c}, M. Sandberg, A. Szepessy} 
\maketitle



\begin{abstract}
In the classical work by Irving and Zwanzig [Irving J.H. and Zwanzig R.W., J. Chem. Phys. 19 (1951), 1173-1180 ] it has been shown that
quantum observables for macroscopic density, momentum and energy
satisfy the conservation laws of fluid dynamics. 
In this work we derive the corresponding classical molecular dynamics limit
by extending Irving and Zwanzig's result to matrix-valued potentials for a general quantum particle system. The matrix formulation provides the  classical limit of the quantum observables in the conservation laws
also in the case where the temperature is large compared to the electron eigenvalue gaps.
The classical limit of the quantum observables in the conservation laws is useful in order to determine the constitutive relations for the stress tensor and the heat flux by molecular dynamics simulations.
The main new steps to obtain the molecular dynamics limit are:
(i)  to approximate the dynamics of quantum observables  accurately by classical dynamics, by diagonalizing the Hamiltonian using  a nonlinear eigenvalue problem,
(ii) to define the local energy density by partitioning a general potential, applying perturbation analysis of the electron eigenvalue problem,
(iii) to determine the molecular dynamics stress tensor and heat flux in the case of several excited electron states, and
(iv) to construct the initial particle phase-space density as a local grand canonical quantum ensemble 
determined by the initial conservation variables.
\end{abstract}

\section{The purpose of the work and the results}
The  macroscopic conservation laws for mass, momentum and energy form the basis of continuum fluid mechanics.
These conservation laws are formulated in terms of the stress tensor and the heat flux.
In order to form a closed system constitutive relations for the stress tensor and the heat flux are used.
Such constitutive relations can be determined approximately from measurements
or from molecular dynamics simulations.   
In both cases one seeks approximations of the stress tensor
and the heat flux as  functions of the density, momentum and energy and their derivatives.
The molecular dynamics formulation requires  derivation of the stress tensor and the heat flux
as functions of the particle dynamics. The derivation of such functional relations is the focus of this work.

The stress tensor and the heat flux were first derived by Irving and Kirkwood, \cite{IK},
from molecular dynamics systems based on interaction with scalar pair potentials and has later been modified by
Noll, \cite{noll}, and Hardy, \cite{hardy}. These formulations have been used frequently to numerically
 determine the constitutive relations, cf. \cite{frenkel}. For instance, 
 the works \cite{li} and \cite{zimmerman2} include comparisons of different methods to
 numerically determine the stress tensor in molecular dynamics simulations.
 
 Already in 1951 Irving and Zwanzig, \cite{IZ}, showed that quantum observables for the density, momentum and energy
 satisfy the conservation laws and derived observables for the stress tensor and the heat flux.
 Since it is only at the quantum level the particle interaction is determined from fundamental principles
 their result provides a solid foundation for the basic conservation laws in continuum mechanics.
 The property that the observables for the density, momentum and energy
 satisfy the conservation laws does not mean that a closed system of conservation laws is derived, since
 the derived stress tensor and the heat flux are not determined as constitutive functions of the macroscopic conservation variables.
 To form a closed system would include the additional step to determine constitutive functions of the conservation variables
 that approximate the 
 data from molecular dynamics or measurements, which is not studied here.
 
 Irving and Zwanzig used a quantum model with the Hamiltonian given by a sum
 of kinetic energy and scalar pair potential energy including all particles,  i.e. both the nuclei and the electrons.
 The aim of this work is to extend the derivation by Irving and Zwanzig to a setting with a matrix-valued
 Hamiltonian consisting of a sum of the kinetic energy of the nuclei (times the identity matrix) and a matrix representing the electron kinetic energy, the electron-electron, 
electron-nuclei, 
and nuclei-nuclei interaction.
 The purpose of having a matrix for the electron part in the Hamiltonian
 is to replace the time evolution for the electrons by the Schr\"odinger electron eigenvalue problem.
  An advantage of including the electron part  as a matrix-valued operator
 is that the classical limit, as the nuclei-electron mass ratio tends to infinity, has been derived rigorously, \cite{teufel} and \cite{KLSS},
 and by knowing the classical limit the system can be simulated by
 ab initio molecular dynamics  for nuclei with the potential generated by the electron eigenvalue problem. 
 For instance, one may ask how the observables of the density, momentum, energy, stress tensor and heat flux
 are effected by the possibility of excited electron states and how these observables should be computed in molecular dynamics simulations. This question is answered in Theorem~\ref{thm:md_limit} by applying
 the classical molecular dynamics limit of quantum observables in \cite{KLSS}. The work   \cite{KLSS}
 is for the setting of constant temperature in the canonical
 ensemble and shows, for example, how the potential is modified also when the difference
 of the excited and ground state electron eigenvalue is not large compared to the temperature.
 
 The time evolution of the conserved quantum observables uses the ingenious observation by Irving and Zwanzig that, for an observable that is a polynomial of the degree at most two in the momentum coordinate, the commutator of the Hamiltonian operator and that quantum observable becomes equal to the Weyl quantization of the Poisson bracket. Combined with the observation that the observables for density, momentum and energy are polynomials of  degree at most two in the momentum coordinate the
 quantum observables therefore satisfy the same conservation laws as in the derivation based on classical particle dynamics
 by Irving and Kirkwood. However, in the case of matrix-valued potentials the commutator of the
 Hamiltonian and the quantum observables for mass, momentum and energy does not reduce to a Poisson bracket since  the  matrix-valued symbols do not commute in general. 
 In this work we show that for a certain diagonalization, based on a nonlinear  eigenvalue problem, 
 these commutators are
 reduced to a quantization of corresponding Poisson brackets. 
 
 To define the energy observable the works \cite{IK,noll,IZ,hardy} use that the potential energy can be split into a sum of potential energies related to each particle as defined by pair interactions.
 In the matrix-valued case considered here the splitting is required for the eigenvalues of the matrix potential which is not a sum of pair potential interactions. Our splitting is instead obtained by using perturbation theory for eigenvalues.
 
 The pair potential property is also used in the works \cite{IK, noll, IZ, hardy} to reduce forcing terms to
  divergence of a stress term. Such reduction has been obtained in \cite{tadmor} for general potentials that are invariant with respect to translation and orthogonal transformations by changing to the coordinates depending on all pair distances.  This change to the pair distance coordinates is also used here.
  
  
  The compressible Euler equations have been derived from classical perturbed Newtonian particle dynamics using the relative entropy  method in \cite{OVY}. The classical Newtonian particle dynamics based on short range pair potential interactions is then weakly perturbed in  two ways: to avoid unbounded velocities the kinetic energy is modified, for instance as relativistic, and to prove ergodicity with respect to Gibbs distributions the Hamiltonian dynamics is perturbed by a weak noise term that vanishes in the macroscopic hydrodynamic limit. A main accomplishment in \cite{OVY} is to show that the density solving the Liouville equation that is initially close to a grand canonical Gibbs measure remains close to a grand canonical Gibbs measure at later time, so that the Gibbs measure determines the compressible Euler equations for all times, as long as the solution to the Euler equations remains smooth. The work \cite{OVY} achieves the mathematically ambitious goal  to derive a closed system of conservation laws from microscopic dynamics, which also requires additional assumptions and restricts to a setting with smooth classical solutions to the Euler equations. The Euler equations includes a pressure term that originates from microscopic particle forces.  The relative entropy method has also been used to derive the compressible Euler equations, with a certain pressure term,   in a scaling limit from a quantum system of fermions  under an assumption of ergodicity of the quantum dynamics with respect to the Gibbs measure, see \cite{NY}.
The objective in our work here is different from \cite{OVY} and \cite{NY}, in particular, we derive microscopic expressions for the stress tensor and the heat flux from a general quantum mechanical setting but we do not address the question of deriving a closed system of conservation laws from quantum mechanics.

We formulate the quantum mechanical model and the conservation laws in Section~\ref{sec_2}. 
In Sections~\ref{sec_classical} and \ref{sec_qm} we review the derivations of the conservation laws
from classical and quantum dynamics, respectively, following the works \cite{IK,noll, hardy,IZ}, although reformulated in order to prepare for the new results in Section~\ref{sec_qm_class}.  These derivations are then used to obtain the matrix-valued extension of the
quantum dynamics in Section~\ref{sec_qm_class}. We derive the main result in Theorem \ref{thm:md_limit}, namely a classical molecular dynamics limit of the quantum observables in the conservation laws, under some assumptions on regularizations:
 the semiclassical analysis result requires $L^2$ bounded symbols, which is not satisfied in the canonical quantum formulation. The theorem therefore assumes that certain regularized $L^2$-bounded symbols depend continuously on the regularization parameters.
In Section~\ref{sec:initial} we present an approach for determining  an initial phase-space density that matches the initial conservation variables locally.
Following \cite{tadmor} we discuss in Section~\ref{sec:derivatives}  the non-uniqueness question for the stress tensor.

In conclusion, the main result in this work is
to formulate the quantum conservation laws using matrix valued symbols and
apply recent techniques from semiclassical analysis to determine 
the molecular dynamics stress tensor and heat flux in the case of several excited electron states. 
A new ingredient in the formulation is also to 
define the local energy density by partitioning a general potential, applying perturbation analysis of the canonical electron eigenvalue problem;
previous  work on molecular dynamics formulations of the conservation laws used empirical pair potentials, \cite{IK,noll,hardy,OVY}.
As in \cite{OVY} we construct the initial particle phase-space density as a local grand canonical quantum ensemble 
determined by the initial conservation variables.

We think that the presented work is  the first ab initio result that determines the stress and the heat flux in a molecular dynamics setting with  several excited states and at any temperature. Therefore we believe that this is a valuable first step for further study. For instance, it would be interesting to extend the result by determining conditions that imply the assumed continuous dependence on the regularization parameters.

 \section{Problem formulation}\label{sec_2}
 
\subsection{The quantum-mechanical model}\label{sec_qm2}
We consider derivation of conservation laws from ab initio dynamics for which the starting point  is the quantum mechanical model consisting of $N$ nuclei
(heavy particles or slow degrees of freedom) 
and $J$ electrons (light particles or fast degrees of freedom). Each particle has a related position coordinate in $\rset^3$ and a discrete spin coordinate. The spin coordinate $\bar\sigma_i$ for each electron takes the value in the set $\{-1/2,1/2\}$, and similarly the spin coordinate $\sigma_i$ for a nucleus can take values in a discrete set $\{-s,-s+1,\dots,s\}$, see \cite{lieb,handbook}. 
The quantum system at time $t$ is then described by a wave function 
$$
\Phi(x^1,\sigma^1,x^2,\sigma^2,\ldots, x^N,\sigma^N,\bar x^1,\bar\sigma^1,\bar x^2,\bar\sigma^2,\ldots, \bar x^J,\bar\sigma^J,t)\in \cset,
$$ 
with 
 nuclei position coordinates $x=(x^1,x^2,\ldots, x^N)\in\rset^{3N}$
and electron position coordinate $\bar x=(\bar x^1,\bar x^2,\ldots, \bar x^J)\in\rset^{3J}$, and spin coordinates $\sigma=(\sigma^1,\ldots,\sigma^N)\in 
\Sigma \equiv \{-s,-s+1,\ldots,s\}^N$ and $\bar\sigma=(\bar\sigma^1,\ldots, \bar\sigma^J)\in \bar\Sigma\equiv\{-1/2,1/2\}^{J}$.
The wave function is required to satisfy the Pauli exclusion principle which implies that
it is anti-symmetric with respect to interchanging electron coordinates, namely
\[
\begin{split}
&\Phi(\ldots,
\bar x^i,\bar\sigma^i,\ldots, \bar x^j,\bar\sigma^j,\ldots ,t)=
-\Phi(\ldots,
\bar x^j,\bar\sigma^j,\ldots, \bar x^i,\bar\sigma^i,\ldots ,t)
\end{split}
\]
and similarly identical fermion nuclei are also anti-symmetric while identical boson nuclei are symmetric with respect to its nucleon coordinates, see \cite{lieb,handbook}.
We note that 
\[
\frac{|\Phi(x,\sigma,\bar x,\bar\sigma,t)|^2}{\sum_{\sigma\in \Sigma}\sum_{ \bar\sigma\in \bar\Sigma}\int_{\rset^{3(N+J)}}
|\Phi(x,\sigma,\bar x,\bar\sigma,t)|^2{\rm d}x{\rm d}\bar x}
\]
is the probability to find the quantum system in  $(x,\sigma,\bar x,\bar\sigma)$ at time $t$.
In the absence of magnetic fields, the wave function depends on the spin coordinates only parametrically since the Hamiltonian does not depend on the spin coordinates.
To simplify the notation we therefore suppress  the spin coordinates 
in the sequel although we include the dependence on the spin in Pauli exclusion principle for the electron wave functions.
We assume the atomic units (a.u.) in which the mass and  charge of the electron are equal to one, and the Planck constant is $\hbar=1$. We denote by $M_n$ the mass of individual nuclei.

The quantum mechanical evolution is described by the Schr\"odinger equation for a wave function
$ \Phi:\rset^{3N}\times\rset^{3J}\times [0,\infty)\rightarrow\cset$ satisfying
\begin{equation}\label{schrodinger}
\IU \, \partial_t \Phi(x,\bar x,t) = \OPER{\HAM} \Phi(x,\bar x,t)\COMMA
\end{equation}
with the Hamiltonian operator 
\begin{equation} \label{fullHamOper}
\OPER{\HAM}= -\sum_{n=1}^{N} \frac{1}{2M_n} \Delta_{x^n} + \OPW(x,\bar x)+  V_b(x) + v_b(x,\bar x)\COMMA
\end{equation} 
where $V_b:\rset^{3N}\to\rset$ and $v_b:\rset^{3N}\times\rset^{3J}$ are given ad hoc external scalar smooth potentials and
$\OPW(x,\bar x)$ is the electronic operator formed by the electron kinetic energy,
electron-electron repulsion, nuclei-nuclei repulsion and electron-nuclei attraction, see \cite{handbook,lebris,lieb},
\begin{equation}\label{v_def}
\begin{split}
\OPW(x,\bar x)   & = -\frac{1}{2} \Delta_{\bar x} + {\nu}(x,\bar x)\\
{\nu}(x,\bar x) & =\sum_{j=1}^J\sum_{k<j} \frac{1}{|\bar x^k-\bar x^j|} 
                      + \sum_{m=1}^N\sum_{n<m} \frac{Z_n Z_m}{|x^n-x^m|}
                       -\sum_{j=1}^J\sum_{n=1}^N \frac{ Z_n}{|x^n-\bar x^j|}\PERIOD
\end{split}
\end{equation}
Here  $Z_n$ denotes the charge of the $n$th nucleus and $\bar x^j\in\rset^3$ the coordinate of the electron $j$. 
The external potentials $V_b:\rset^{3N}\to\rset$ and $v_b:\rset^{3N}\times\rset^{3J}\to \rset$ 
make the system confined and do not perturb the system far away from the boundary by 
the assumptions 
\begin{equation}\label{vbdry}
\begin{split}
& v_b(x,\bar x) :=  \sum_{j=1}^J  h(|\bar x^j-x^0|), \\
& h(y)\to\infty\quad\mbox{ as $y\to\infty$},\\
& h(y)=0\quad \mbox{ for $|y|\le C$,}\\
& V_b(x)\to\infty\quad\mbox{ as $|x|\to\infty$},\\
& V_b(x)=0\quad \mbox{ for $|x|\le C$,}
\end{split}
\end{equation}
with a large constant $C$, where $x^0:=\sum_{n=1}^N x^n/N$
and $h:\rset\to \rset$ is
a given smooth function. 

We now show how the summation over all particles in the definition of the potential operator \eqref{v_def} can be rearranged into a sum over contributions from each nucleus
\begin{equation}\label{V_nsum}
\OPW = \sum_{n=1}^N \OPW^n.
\end{equation}
We partition the electron index set $\mathcal{E}\equiv\{1,\dots,J\}$ into disjoints subsets $\mathcal{E}_n$, $n=1,\dots,N$ such that $\bigcup_{n=1}^N \mathcal{E}_n = \mathcal{E}$ and the number of elements in $\mathcal{E}_n$ is equal to $Z_n$, i.e., $\# \mathcal{E}_n = Z_n$. For the ease of exposition we assume charge neutrality of the electron-nuclei system.
Using this partitioning we define
\begin{equation}\label{V_n}
\OPW^n(x,\bar x) =
-\frac{1}{2} \sum_{k\in \mathcal E_n} \Delta_{\bar x^k} 
+ \frac{1}{2}\sum_{\ell\in \mathcal E_n}\sum_{k\ne\ell} \frac{1}{|\bar x^k-\bar x^\ell|}  +\frac{1}{2} \sum_{m\neq n} \frac{Z_nZ_m}{|x^n-x^m|}
-\sum_{k=1}^J \frac{ Z_n}{|x^n-\bar x^k|}\COMMA 
\end{equation}
where the terms including the
sums in $\mathcal{E}_n$ correspond to the electron kinetic energy and electron-electron repulsion for electrons associated with the  nucleus $n$.

\subsection{ The electronic operator}\label{sub_sub_electron}

The electron eigenvalue problem takes the form
\begin{equation}\label{eigen_v*}
(\OPW+v_b)\Psi_k=\tLm_k\Psi_k\COMMA\;\;\; 
\end{equation}
with each function $\Psi_k$ in the anti-symmetric wave function subspace of $L^2$, based on $\bar x$ and the electron spin coordinates.
Since $\OPW+v_b$ depends on the nuclei coordinate $x$ only parametrically, the eigenfunctions and eigenvalues will depend on $x$. We will see that the eigenvalues $\tLm_k, \ k=1,\ldots$ will be the potentials for the nuclei dynamics which determine the molecular dynamics. The work \cite{IK,IZ,noll,hardy} use the explicit pair interactions of their potentials to define the energy related to each particle and to treat the conservation of total momentum. Our ab initio potentials $\tLm_k$ are not given by pair interactions.
To derive the conservation laws from the quantum mechanical formulation \eqref{schrodinger}, we will in particular use two properties of the electron eigenvalue problem \eqref{eigen_v}, namely its invariance with respect to translations and rotations in $\rset^3$
and its construction by pair interactions.  

\subsubsection{Rotational invariance}\label{rot_inv} We note that $\OPW+v_b$ is invariant with respect to the affine transformations defined by 
\[
(x^1,\ldots,x^N,\bar x^1,\ldots, \bar x^J) \mapsto 
(Qx^1+\alpha ,\ldots, Qx^N+\alpha,Q\bar x^1+\alpha, \ldots,Q\bar x^J+\alpha)\COMMA
\]
where $Q\in O(3)$ is an orthogonal transformation of $\rset^3$ and $\alpha\in \rset^3$ a translation, i.e.,
\[
(\OPW+v_b)(x,\bar x)=(\OPW+v_b)(Qx^1+\alpha ,\ldots, Qx^N+\alpha,Q\bar x^1+\alpha, \ldots,Q\bar x^J+\alpha) \PERIOD
\] 
Therefore also the eigenvalues $\tLm_k$ and eigenfunctions are invariant with respect to such translations and orthogonal transformations. 

\subsubsection{ Partition of the energy}\label{en_part} 
We define the potential energy related to the nucleus $n$ as 
\begin{equation}\label{lambda_n_def}
\tLm_k^n(x)= \SCPRODE{\Psi_k(x)}{\Big(\OPW^n+\frac{v_b}{N}\Big)(x)\Psi_k(x)}\PERIOD
\end{equation}
Here we denote by $\SCPRODE{u}{w} =\sum_{\bar\sigma\in\bar\Sigma}\int_{\rset^{3J}} u^*(\bar x,\bar\sigma) w(\bar x,\bar\sigma)\,\Rd \bar x$ the scalar product on the Hilbert space $L^2(\rset^{3J}\times \bar\Sigma)$ of electron states.
This definition can be motivated by standard perturbation analysis, see \cite{kato}: A small perturbation $ w$ of the potential $\OPW +v_b$
yields to leading order that the $k$th eigenvalue $\tLm_k=\SCPRODE{\Psi_k}{(\OPW+v_b)\Psi_k}$  is perturbed to 
$\SCPROD{\Psi_k}{\big( \OPW+ v_b+ w\big)\Psi_k}$.
The energy related to the particle $n$, for eigenvalue $k$, can therefore be viewed as the
difference of  $\tLm_k$ and the $k$th eigenvalue for the perturbed potential $\OPW+v_b-(\OPW^n+v_b/N)$,
which corresponds to removing particle $n$ and its electrons and the fraction $1/N$ of the external potential $v_b$. 
This perturbed eigenvalue is, to the leading order,
\[
\SCPRODE{\Psi_k}{\big( \OPW+v_b-(\OPW^n+v_b/N)\big)\Psi_k\Big)}\, ,
\] 
which leads to the definition in \eqref{lambda_n_def}.
Property \eqref{V_n} and \eqref{lambda_n_def} imply that for all $k$
\begin{equation}
\tLm_k = \sum_{n=1}^N \tLm_k^n \PERIOD \label{A3} 
\end{equation}

%
In the introductory Section \ref{sec_classical} we 
will consider only one energy surface, often the ground state $\tLm_1(x)$, and therefore we drop the index $k$ there, and use the notation $\lambda:=\tLm_k$ for a given fixed $k$. In Section \ref{sec_qm_class} we use all electron eigenvalues $\tLm_k, \ k=1,\ldots d$, to establish molecular dynamics representations of the stress tensor and the heat flux.

\subsubsection{Reduction to finite dimensional electron space and regularization of the electron operator} 

To obtain the classical limits in Section~\ref{sec_qm_class} 
we first regularize the electron operator $\OPW+v_b$ 
by replacing all Coulomb terms $1/|z|$ in $\nu$ by $1/\sqrt{|z|^2+\delta_a}$, with a positive constant $\delta_a>0$. We note that this regularization preserves the translation and rotation invariance. Then we introduce a finite dimensional
approximation $\tVd$ of $\OPW+ v_b$ in the anti-symmetric electronic wave function space  
(including also the spin coordinates)
by applying a projection onto the finite-dimensional subspace of
$\LSPEL$ spanned by the eigenfunctions 
$\{\Psi_1,\dots,\Psi_d\}$ of the regularized Hamiltonian $\OPW+v_b$, with the corresponding eigenvalues $\{\tLm_1,\ldots,\tLm_d\}$,
as 
\begin{equation}\label{V_definitionen}
\begin{split}
{ \tVd}\psi &=\sum_{k=1}^d\tLm_k\Psi_k\SCPRODE{\Psi_k}{\psi}\COMMA\\
{ \tVd}^n\psi &=\sum_{k=1}^d\tLm_k^n\Psi_k\SCPRODE{\Psi_k}{\psi}\COMMA
\end{split}
\end{equation}
for any wave function $\psi\in \LSPEL$. 
Since we have $v_b(x,\bar x)\to\infty$ as $|\bar x|\to\infty$ for given $x$ and the potential part in $\OPW+v_b$ is locally integrable with respect to $\bar x$,
the spectrum of $\OPW +v_b$ is discrete, see \cite{mazja}.


The motivation to introduce the reduction \eqref{V_definitionen}  to a 
smooth $d\times d$ matrix with distinct eigenvalues is that we then have a precise result of the approximation of quantum observables by molecular dynamics, in particular Theorem \ref{gibbs_corr_thm} proved in \cite{KLSS}. 
Thus $\OPW+v_b$ is approximated by a Hermitian matrix-valued operator $V:\rset^{3N}\to \cset^{d\times d}$ by \eqref{V_definitionen}.
We  assume that the eigenvalues 
$\lambda_1(x),\lambda_2(x),\ldots, \lambda_d(x)$ 
of this matrix-valued potential $V(x)$, i.e., solutions
\begin{equation}\label{eigen_v}
V(x)\psi_k(x)=\lambda_k(x)\psi_k(x)\COMMA\;\;\;\mbox{for all $x\in\rset^{3N}$,}
\end{equation}
satisfy
\begin{align}
 &\lambda_1(x)<\lambda_2(x)<\ldots<\lambda_d(x)\COMMA \label{A1}\\
 &\lambda_1(x)+ {V_b}(x)\rightarrow\infty \;\;\;\mbox{ as $|x|\rightarrow \infty$,} \label{A2}\\ 
 &\lambda_d(x)\to \infty 
 \;\;\;\mbox{ as $d\to\infty$},\label{A21}\\
 &\{\lambda_k\}_{k=1}^d  \mbox{ depend continuously on $\delta_a$ as $\delta_a\to 0+$}\, .\label{A22}
\end{align}
The first assumption is in order to have differentiable eigenvectors, the second condition implies that the system is confined, with respect to the nuclei, and the third condition is used to have a discrete spectrum of $\OPW+v_b$ and ensure that $V$ is a consistent approximation of $\OPW+v_b$.

In general, nuclei positions $x$ where electron eigenvalues coincide, e.g. points $x$
where $\tLm_1(x)=\tLm_2(x)$, form a co-dimension two set in $\rset^{3N}$.
Such points include so-called conical intersections which
are difficult to handle in the classical limit and they are not included in Theorem~\ref{gibbs_corr_thm}.
Section~\ref{sec_conical} describes how an extension of Theorem~\ref{gibbs_corr_thm} to include conical intersection could be possible.

\subsection{The conservation laws}\label{sec_cons}

The conservation laws or balance laws for mass, momentum and energy, based on the density 
$\rho:\rset^3\times[0,\infty)\rightarrow[0,\infty)$, velocity $u:\rset^3\times [0,\infty)\rightarrow\rset^3$ 
and energy density 
$E:\rset^3\times[0,\infty)\rightarrow\rset$, take the form
\begin{equation}\label{conservation_law}
\begin{split}
\partial_t\rho(y,t) + \sum_{\ell=1}^3\partial_{y_\ell}\big(\rho(y,t)u_\ell(y,t)\big) &=0\, ,\\
\partial_t\big(\rho(y,t)u_j(y,t)\big) + \sum_{\ell=1}^3\partial_{y_\ell}\big(\rho(y,t)u_j(y,t)u_\ell(y,t)- \sigma_{\ell j}(y,t)\big) &=F_j(y,t)\, ,\\
\partial_t E(y,t) + \sum_{\ell=1}^3\partial_{y_\ell}\big( E(y,t)u_\ell(y,t) + q_\ell(y,t)- \sum_{j=1}^3\sigma_{\ell j}(y,t)u_j(y,t) \big) &= P(y,t)\, ,\\
\end{split}
\end{equation}
where $\sigma_{\ell j}:\rset^3\times[0,\infty) \rightarrow\rset$ is the $\ell j$-component of the $3\times 3$ stress tensor, $q_\ell:\rset^3\times[0,\infty)\rightarrow\rset$ is the $\ell$-th component of the heat flux,
$F:\rset^3\times[0,\infty)\to\rset^3$ is an external force and $P:\rset^3\times[0,\infty)\to\rset$ is an external energy source. The purpose of this work
is to derive these conservation/balance laws from microscopic dynamical systems.  First we consider classical systems and then quantum systems.

%
%
\section{The conservation laws derived from classical particle dynamics}\label{sec_classical}

In this section we consider a system of $N$ classical particles, where each particle has mass $M_n$, position coordinate $x^n$ and momentum coordinate $p^n$, $n=1,\ldots, N$. We use the notation $x=(x^1,x^2,x^3,\ldots, x^N)$ and $p=(p^1,p^2,p^3,\ldots, p^N)$. The
position coordinates $x:[0,\infty)\rightarrow \rset^{3N}$ and momentum coordinates 
$p:[0,\infty)\rightarrow \rset^{3N}$ satisfy the classical equations of motion given by the Hamiltonian 
\begin{equation}\label{classical_Hamiltonian}
    H(x,p) = \sum_{n=1}^N \frac{1}{2M_n}|p^n|^2 + \lambda(x)+V_b(x)\COMMA
\end{equation} 
where $\lambda:\rset^{3N} \to \rset$ is a given interaction potential which we relate to the potential energy surface, i.e. an eigenvalue $\lambda_k(x)$ in \eqref{eigen_v}, in the next section and $V_b:\rset^3\to\rset$ is the external potential. 
Thus the evolution of the system is given by the solution
of Newtonian dynamics
\begin{equation*}
\begin{split}
\dot x^n_t &= \frac{1}{M_n} p^n_t\COMMA\;\;\mbox{$n=1,\dots,N$}\\
\dot p_t &= -\nabla\lambda(x_t)-\nabla V_b(x_t)\COMMA 
\end{split}
\end{equation*}
where $x^n_t\in \rset^3$ and $p^n_t\in \rset^3$ is the position and momentum, respectively,  of the particle $n$
at time $t$.  
 The given potential $\lambda:\rset^{3N}\rightarrow\rset$
is assumed to be invariant 
under the Euclidean group of transformations of $\rset^3$,
i.e. 
$\lambda(x^1,\ldots,x^N)=\lambda(Qx^1+\alpha, \ldots, Qx^N+\alpha)$ for any orthogonal $3\times 3$ matrix $Q$
and any translation $\alpha\in\rset^3$.
The following lemma will be used to represent the potential
$\lambda$ as  a function of pairwise distances between particles
rather than the positions of each particle. 
It shows that there is a transformation in the Euclidean group, which consists of isometries, that maps all elements of one point set to another point set, provided the distances between points in both sets coincide.
\begin{lemma}\label{thm:rottransl}
If the two sets of points $\{x^i\}_{i=1}^N$  and $\{y^i\}_{i=1}^N$,
where $x^i, y^i \in \rset^3$, satisfy $r^{ij}=|x^i-x^j|=|y^i-y^j|$ for
$1\leq i,j\leq N$, then there exist an orthogonal matrix
$Q\in\rset^{3\times 3}$ and a translation vector $\alpha\in\rset^3$
such that $x^i=Qy^i+\alpha$ for $1\leq i \leq N$.
\end{lemma}
\begin{proof}
Let $\bar x^i:=x^i-x^1, \bar y^i=y^i-y^1$, for $1\leq i \leq N$. If
$\bar x^i=\bar y^i=0$, for all $1\leq i \leq N$, then clearly the
claim in the theorem is true. If not, let $i_1$ be an index such that
$\bar x^{i_1}\neq 0$ (which also implies that $\bar y^{i_1}\neq
0$). Let $Q_1,Q_2\in\rset^{3\times 3}$ be two orthogonal matrices such
that $Q_1\bar x^{i_1}$ and $Q_2\bar y^{i_1}$ both lie on the first
positive coordinate axis. Then clearly $Q_1\bar x^{i_1}=Q_2\bar
y^{i_1}$. 

Define $\bar{\bar x}^i:=Q_1\bar x^i$ and $\bar{\bar y}^i:=Q_2\bar y^i$
for all $1 \leq i \leq N$.
If all $\bar{\bar x}^i$ and $\bar{\bar y}^i$ lie on the first coordinate axis then
$\bar{\bar x}^i=\bar{\bar y}^i$, for $1\leq i\leq N$, since every
$\bar{\bar x}^i$ and $\bar{\bar y}^i$
have the same distance to $\bar{\bar x}^1$ in the origin, and
$\bar{\bar x}^{i_1}$. Assume now that there exists an index $i_2$ such that $\bar{\bar
x}^{i_2}$ does not lie on the first coordinate axis. Since $\bar{\bar
  x}^{i_2}$ and $\bar{\bar y}^{i_2}$ have the same distance to $\bar{\bar x}^1$  and
$\bar{\bar x}^{i_1}$, also $\bar{\bar y}^{i_2}$ does not lie on the
first coordinate axis. Let $Q_3,Q_4\in\rset^{3\times 3}$ be two
orthogonal matrices that are rotations around the first coordinate
axis such that $Q_3 \bar{\bar x}^{i_2}$ and $Q_4 \bar{\bar y}^{i_2}$
are both in the ``positive $xy$-plane'', i.e.\ given as $(a,b,0)$ for
$b>0$. This makes $Q_3 \bar{\bar x}^{i_2}=Q_4 \bar{\bar y}^{i_2}$
since the points are on the same distance to $\bar{\bar x}^1$ and
$\bar{\bar x}^{i_1}$.

Define $\bar{\bar{\bar x}}^i:=Q_3\bar{\bar x}^i$ and $\bar{\bar{\bar
    y}}^i:=Q_4\bar{\bar y}^i$ for all $1\leq i \leq N$. Since the points $\bar{\bar{\bar x}}^i$
and  $\bar{\bar{\bar y}}^i$ have the same distance to the points
$\bar{\bar{\bar x}}^1$, $\bar{\bar{\bar x}}^{i_1}$, and
$\bar{\bar{\bar x}}^{i_2}$, that all lie in the plane spanned by the
first two coordinate directions, but not all of them on a straight
line, we must either have that $\bar{\bar{\bar x}}^i=\bar{\bar{\bar
    y}}^i$ or $\bar{\bar{\bar x}}^i=Q\bar{\bar{\bar
    y}}^i$, for the reflection in the $xy$-plane
$Q=\bigl(\begin{smallmatrix} 1&0&0\\ 0&1&0 \\ 0&0&
  -1\end{smallmatrix}\bigr)$. There cannot be two points
  $\bar{\bar{\bar x}}^i$ and   $\bar{\bar{\bar x}}^j$ that do not lie
  in the $xy$-plane and satisfy   $\bar{\bar{\bar x}}^i=\bar{\bar{\bar
      y}}^i$ and $\bar{\bar{\bar x}}^j=Q\bar{\bar{\bar y}}^j$, since
  then $\bar{\bar{\bar x}}^j$ and $\bar{\bar{\bar y}}^j$ would be on
  different distance from $\bar{\bar{\bar x}}^i=\bar{\bar{\bar
      y}}^i$.

Hence either $\bar{\bar{\bar x}}^i=\bar{\bar{\bar
      y}}^i$ for all $1 \leq i \leq N$, or $\bar{\bar{\bar x}}^i=Q\bar{\bar{\bar
      y}}^i$ for all $1 \leq i \leq N$. Since  $\bar{\bar{\bar
    x}}^i$ are obtained from $x^i$ by the same set of  translations and
multiplications by orthogonal matrices for all $1\leq i \leq N$, and
likewise for $\bar{\bar{\bar y}}^i$, the  proof is complete.

\end{proof}
 
To handle conservation of total momentum we will use Newtons third law for pair interactions
and we follow the construction in \cite{tadmor} to determine pair interactions in  a general potential
that is invariant with respect to translations and orthogonal transformations in $\rset^3$:
 knowing all $N(N-1)/2$ pair distances 
$r:=(r^{12},r^{13},\ldots,r^{N-1 N}):=(|x^1-x^2|, |x^1-x^3|,\ldots
 ,|x^{N-1}-x^N|)$  
determines $x$ up to a translation and orthogonal transformation in
$\rset^3$ and since $\lambda(x)$ remains the same for such
translations and orthogonal transformations the potential is
determined by all pair distances, i.e. 
\begin{equation}\label{eq:tildelambda}
\lambda(x)=:\tilde
\lambda(r(x)).
\end{equation}  
We will use the partial derivatives
$\partial_{r^{jk}}\tilde\lambda(r^{12},r^{13},\ldots,r^{N-1 N})$\, .
Not all $r\in\rset^{N(N-1)/2}$ correspond
to particle positions $x\in \rset^{3N}$ and
there are $N(N-1)/2$  partial derivatives $\partial_{r^{jk}}\tilde\lambda(r)$
while the gradient $\nabla\lambda(x)$ only has $3N$ components.
Therefore the partial derivatives $\partial_{r^{jk}}\tilde\lambda(r)$
are not uniquely defined by $\nabla\lambda(x)$.
Section \ref{sec:derivatives} shows how to determine $\partial_{r^{jk}}\tilde\lambda(r)$.

To define the observables for density, momentum and energy and their dependence on the space coordinate $y\in\rset^3$ we use a non-negative smooth  mollifier $\eta:\rset^3\rightarrow\rset$, 
$\eta\in C^\infty(\rset^3)$, with a compact support,
 satisfying
\[
\begin{split}
\int_{\rset^3}\eta(y) \Rd y &=1\, ,\\
\eta(y) &\geq 0\COMMA\;\;\; \eta(y) = \eta(-y)\COMMA\;\; \quad \mbox{for all $y\in\rset^3$,}\\
\eta(y) &= 0 \, ,\quad \mbox{ for } |y|>\epsilon\, .
\end{split}
\]
The macroscopic density $\rho: \rset^3\times [0,\infty)\rightarrow\rset$ is defined by the particle system as
\begin{equation*}
\begin{split}
\rho(y,t) &= \int_{\rset^{6N}} \sum_n M_n\,\eta(y-x_t^n) f(x_0,p_0) \Rd x_0\Rd p_0\, ,
\end{split}
\end{equation*}
where $x_t^n$ is a function of the initial condition $(x_0,p_0)$, and  $f:\rset^{6N}\rightarrow [0,\infty) $
is a given initial particle distribution function normalized so that $\int_{\rset^{6N}}f(x_0,p_0)\Rd x_0\Rd p_0=1$. 
Irving and Kirkwood, \cite{IK},
use this definition with $\eta$ equal to a point mass and a general initial distribution $f$.
 Noll, \cite{noll}, formulates the integration with respect to point masses in terms of the one-point and two-point density correlations functions instead and provides precise conditions for the validity of the derivation. Hardy, \cite{hardy},  uses  the mollifier $\eta$ but not
the integration over the initial particle distribution.

\subsection{The conservation of mass} Let $z_0=(x_0,p_0)$ denote the phase-space coordinate in $\rset^{6N}$ and $x\cdot y=\sum_{i=1}^3 x^i y^i$ the Euclidean scalar product in $\rset^3$.  Differentiation of the density implies
\begin{equation}\label{continuityEq}
\begin{split}
\partial_t \rho(y,t) &= 
-\int_{\rset^{6N}} \sum_{n=1}^N M_n\dot x_t^n\cdot\nabla\eta(y-x^n_t)  f(z_0)\Rd z_0\\
&= -\int_{\rset^{6N}} \sum_{n=1}^N p_t^n\cdot\nabla\eta(y-x^n_t)  f(z_0)\Rd z_0\\
\end{split}
\end{equation}
and by defining the velocity $u:\rset^3\times [0,\infty)\rightarrow \rset^3$ as
\begin{equation}\label{u_def}
\rho(y,t) u(y,t):=  \int_{\rset^{6N}} \sum_{n=1}^N \eta(y-x^n_t) p_t^n f(z_0)\Rd z_0
\end{equation}
we obtain the conservation law for the mass
\[
\partial_t\rho(y,t) + \sum_{k=1}^3\partial_{y_k}\big(\rho(y,t)u_k(y,t)\big) =0\, .
\]

\subsection{The conservation of momentum}
Differentiation of the momentum yields
\[
\begin{split}
\partial_t \big(\rho(y,t)u(y,t)\big) &= 
-\int_{\rset^{6N}} \sum_{n=1}^N M_n^{-1}p^n_t\cdot\nabla\eta(y-x^n_t) p_t^n  f(z_0)\Rd z_0\\
&\quad -\int_{\rset^{6N}} \sum_{n=1}^N \eta(y-x^n_t) \nabla_{x^n}\lambda(x_t) f(z_0)\Rd z_0\\
&\quad -\int_{\rset^{6N}} \sum_{n=1}^N \eta(y-x^n_t) \nabla_{x^n}V_b(x_t) f(z_0)\Rd z_0\, .\\
\end{split}
\]
In order to write the second term as a divergence term we follow Noll's, \cite{noll}, and Hardy's method, \cite{hardy}, based
on identifying gradients with respect to pair distances and converting the difference
in $\eta$ at the  the corresponding point to a gradient term:  the
combination of the pair distance derivative, using the definition of
$\tilde\lambda$ in \eqref{eq:tildelambda},
\begin{equation}\label{chain_r}
\begin{split}
\sum_{n=1}^N \eta(y-x^n)\nabla_{x^n}\lambda 
&= \sum_n \sum_{j<k} \eta(y-x^n) \partial_{r^{jk}}\tilde\lambda(r)\,
\nabla_{x^n}(|x^j-x^k|)\\
&= \sum_{n<k}\big(\eta(y-x^n)-\eta(y-x^k)\big) 
\partial_{r^{nk}}\tilde\lambda(r)\, \nabla_{x^n}(|x^n-x^k|)\\
\end{split}
\end{equation}
and the difference at the corresponding points
\begin{equation}\label{eta_diff}
\begin{split}
\eta(y-x^n)-\eta(y-x^k) & = \int_0^1 \frac{\Rd }{\Rd s}\eta\big(y-sx^n-(1-s)x^k\big) \Rd s\\
&=  \int_0^1 (x^k-x^n)\cdot \nabla \eta\big(y-sx^n-(1-s)x^k\big) \Rd s\\
&= -{\mathrm{div}}_y  \int_0^1 \eta\big(y-sx^n-(1-s)x^k\big) (x^n-x^k)\,\Rd s\, \\
\end{split}
\end{equation}
shows that
\[
\begin{split}
\sum_n \eta(y-x^n)\nabla_{x^n}\lambda &=  -{\mathrm{div}_y}
\Big(\sum_{n<k} \int_0^1 \eta(y-sx^n-(1-s)x^k) (x^n-x^k)\,\Rd s\Big)\\
&\qquad \times\partial_{r^{nk}}\tilde\lambda(r) \,\nabla_{x^n}|x^n-x^k|\, .\\
\end{split}
\]
We conclude that the following conservation law for the momentum holds
\begin{equation}\label{mom_class}
\begin{split}
\partial_t \big(\rho(y,t)u_j(y,t)\big) &= 
-\sum_{\ell=1}^3 \partial_{y_\ell} \int_{\rset^{6N}} \sum_{n=1}^N M^{-1}_n \eta(y-x^n_t) p_j^n(t) p^n_\ell(t) f(z_0)\Rd z_0\\
&\quad 
+\sum_{\ell=1}^3 \partial_{y_\ell} \int_{\rset^{6N}} 
\sum_{n<k}  \int_0^1 \eta(y-sx_t^n-(1-s)x_t^k) (x_\ell^n(t)-x_\ell^k(t)) \Rd s\\
&\qquad \times\partial_{r^{nk}}\tilde\lambda \,\partial_{x_j^n} (|x_t^n-x_t^k|) f(z_0)\,\Rd z_0\, .\\
&\quad -\int_{\rset^{6N}} \sum_{n=1}^N \eta(y-x^n_t) \nabla_{x^n} V_b(x_t) f(z_0)\Rd z_0\PERIOD\\
\end{split}
\end{equation}
To write the conservation law for momentum in the form \eqref{conservation_law} we follow the steps in \cite{noll}.
Let 
\[
v^n_t:=\frac{p_t^n}{M_n} -u(y,t)
\]
where  by  \eqref{u_def} 
\[
u(y,t)= \frac{ \int_{\rset^{6N}} \sum_{n=1}^N \eta(y-x^n_t) p_t^n f(z_0)\Rd z_0}{\int_{\rset^{6N}} \sum_{n=1}^N M_n\eta(y-x^n_t)  f(z_0)\Rd z_0}\, .
\]
Definition \eqref{u_def} 
implies
\[
\int_{\rset^{6N}} \sum_{n=1}^NM_n\eta(y-x_t^n)v_t^nf(z_0)\Rd z_0=0\, 
\]
so that the first integral in \eqref{mom_class} satisfies 
\[
\begin{split}
& \int_{\rset^{6N}}\sum_{n=1}^N M^{-1}_n \eta(y-x^n_t) p_j^n(t) p^n_\ell(t)f(z_0)\Rd z_0\\
&=
 \int_{\rset^{6N}} \sum_{n=1}^N M_n \eta(y-x^n_t) \big(v_j^n(t) v^n_\ell(t) + u_j(t) v^n_\ell(t) + v^n_j(t)u_\ell(t) + u_j(t) u_\ell(t)\big)f(z_0)\Rd z_0\\
  &= \int_{\rset^{6N}}\sum_{n=1}^N M_n \eta(y-x^n_t) \big(v_j^n(t) v^n_\ell(t)  + u_j(t) u_\ell(t)\big)f(z_0)\Rd z_0\, .\\
  \end{split}
  \] 
 The conservation law for the momentum can therefore be formulated as
 \[
 \begin{split}
 &\partial_t\big(\rho(y,t)u_j(y,t)\big) + \sum_{\ell=1}^3\partial_{y_\ell}\big(\rho(y,t)u_j(y,t)u_\ell(y,t)- \sigma_{\ell j}(y,t)\big)=F_j(y,t)\, , \\
 \end{split}
 \]
 where
 \begin{equation}\label{sigma_def}
 \begin{split}
 \sigma_{\ell j}(y,t) &:=  \int_{\rset^{6N}} \tilde\sigma_{\ell j}(x_t,p_t;y,t) f(x_0,p_0) \Rd x_0 \Rd p_0\, ,\\
  \tilde\sigma_{\ell j}(x,p;y,t)&:=- 
 \sum_{n=1}^N M_n \eta(y-x^n) v_j^n v^n_\ell \\
&\quad 
+ \sum_{n<k}  \int_0^1 \eta(y-sx^n-(1-s)x^k) (x_\ell^n-x_\ell^k)\Rd s\\
&\qquad \times\partial_{r^{nk}} \tilde\lambda \,\partial_{x_j^n}(|x^n-x^k|)\, ,\\
\end{split}
\end{equation}
defines the stress tensor using $v^n=M_n^{-1}p^n-u(y,t)$,
and
\begin{equation}\label{F_def}
F(y,t):=-\int_{\rset^{6N}} \sum_{n=1}^N \eta(y-x^n_t) \nabla_{x^n} V_b(x_t) f(z_0)\Rd z_0
\end{equation}
defines the macroscopic external force.
  
\subsection{The conservation of energy}
In order to define the energy density we need to define the potential energy related to each particle.
In the case of pair potentials this is straight forward by summing the pair potentials including the particle, as
in \cite{IK,noll,hardy}.  Since we have a more general potential, which does not have to be a sum of pair potentials,  this step requires a new construction: here we use the potential energy $\lambda^n$, related to the particle $n$, introduced in \eqref{lambda_n_def} and \eqref{A3}. 

We define  the energy density  $E:\rset^3\times [0,\infty)\rightarrow\rset$ by
\[
\begin{split}
E(y,t) &= 
\int_{\rset^{6N}} \sum_n \eta(y- x^n_t) \big(\frac{|p_t^n|^2}{2M_n} + \lambda^n(x_t)\big) f(x_0,p_0) \Rd x_0\Rd p_0
\end{split}
\]
and differentiate it to obtain
\begin{equation}\label{E_t}
\begin{split}
\partial_t E_t(y,t) &= -\int_{\rset^{6N}}\sum_n M_n^{-1}p^n_t\cdot \nabla\eta(y-x^n_t)
 \big(\frac{|p^n|^2}{2M_n} + \lambda^n(x_t)\big) f(x_0,p_0) \Rd x_0\Rd p_0\\
&\quad -\int_{\rset^{6N}}\sum_n M_n^{-1}\eta(y-x^n_t)p_t^n
\cdot\nabla_{x^n}\big(\lambda(x_t)+V_b(x_t)\big) f(x_0,p_0) \Rd x_0\Rd p_0\\
&\quad +\int_{\rset^{6N}}\sum_{n,m}M_m^{-1} \eta(y-x^n_t)p_t^m\cdot\nabla_{x^m}\lambda^n(x_t) 
f(x_0,p_0) \Rd x_0\Rd p_0\, .\\
\end{split}
\end{equation}
We have
\[
\begin{split}
\sum_n M_n^{-1}\eta(y-x^n_t) p_t^n\cdot\nabla_{x^n}\lambda(x_t) &=
\sum_{n,m} M_n^{-1}\eta(y-x^n_t) p_t^n\cdot\nabla_{x^n}\lambda^m(x_t)\\
 &= \sum_{n,m} M_m^{-1}\eta(y-x^m_t) p_t^m\cdot\nabla_{x^m}\lambda^n(x_t)\, ,
 \end{split}
 \]
 so the right hand side in \eqref{E_t} becomes
 \[
 \begin{split}
 \partial_t E(y,t) &= -\sum_{\ell=1}^3 \partial_{y_\ell} 
 \int_{\rset^{6N}}\sum_nM_n^{-1} p_\ell^n(t) \eta(y-x^n_t)
 \big(\frac{|p^n|^2}{2M_n} + \lambda^n(x_t)\big) f(x_0,p_0) \Rd x_0\Rd p_0\\
 &\quad + \int_{\rset^{6N}}\sum_{n,m} \big(\eta(y-x^n_t)- \eta(y-x^m_t)\big)
 (M_m^{-1}p_t^m)\cdot\nabla_{x^m}\lambda^n(x_t) 
f(x_0,p_0) \Rd x_0\Rd p_0\\
&\quad -\int_{\rset^{6N}}\sum_n M_n^{-1}\eta(y-x^n_t)p_t^n
\cdot\nabla_{x^n} V_b(x_t) f(x_0,p_0) \Rd x_0\Rd p_0\, .\\
  \end{split}
  \]
By using \eqref{eta_diff} we obtain the conservation law for the energy
 \begin{equation}\label{en_class}
 \begin{split}
& \partial_t E(y,t) \\
&= -\sum_{\ell=1}^3 \partial_{y_\ell} 
 \int_{\rset^{6N}}\sum_n M_n^{-1}p_\ell^n(t) \eta(y-x^n_t)
 \big(\frac{|p^n|^2}{2M_n} + \lambda^n(x_t)\big) f(x_0,p_0) \Rd x_0\Rd p_0\\
 &\quad - \sum_{\ell=1}^3 \partial_{y_\ell}
 \int_{\rset^{6N}}\sum_{n,m} \int_0^1 \eta\big(y-x^m_t+s(x^m_t-x^n_t)\big) \Rd s\big(x^n_\ell(t)-x^m_\ell(t)\big)\\
&\qquad\times (M_m^{-1}p_t^m)\cdot\nabla_{x^m}\lambda^n(x_t) 
f(x_0,p_0) \Rd x_0\Rd p_0\\
&\quad -\int_{\rset^{6N}}\sum_n M_n^{-1}\eta(y-x^n_t)p_t^n
\cdot\nabla_{x^n} V_b(x_t) f(x_0,p_0) \Rd x_0\Rd p_0\, .\\
  \end{split}
  \end{equation}
In order to write the energy conservation in the standard form \eqref{conservation_law}
we use
\[
\begin{split}
\sum_{n=1}^N\eta(y-x^n)(u_\ell+v_\ell^n)\big(\frac{|p^n|^2}{2M_n}+ \lambda^n(x)\big) 
&=\sum_{n=1}^N\eta(y-x^n)u_\ell\big(\frac{|p^n|^2}{2M_n}+ \lambda^n(x)\big)\\
&\quad+ \sum_{n=1}^N\eta(y-x^n)v_\ell^n\big(\frac{M_n|v^n|^2}{2} +\lambda^n(x)\big)\\
&\quad+\sum_{n=1}^N\eta(y-x^n)M_n v_\ell^n\frac{ |u|^2}{2}\\
&\quad+\sum_{n=1}^N\eta(y-x^n)M_n v_\ell^n u\cdot v^n\, .\\
\end{split}
\]
The first term in the right hand side becomes $E(y,t)u_\ell(y,t)$
under the integration in \eqref{en_class}, and the last term is part of
$\sum_j\sigma_{\ell j}(y,t)u_j(y,t)$. The term including the factor $v^n_\ell |u|^2$ vanishes  upon integration over the initial distribution
due to the definition \eqref{u_def} and  the second term is included in the heat flux.
The conservation of energy can therefore be written
\[
\partial_t E(y,t) + \sum_{\ell=1}^3\partial_{y_\ell}\big( E(y,t)u_\ell(y,t) + q_\ell(y,t)- \sum_j\sigma_{\ell j}(y,t)u_j(y,t) \big) 
=P(y,t)\]
where the heat flux is defined as
\begin{equation}\label{heat_def}
\begin{split}
q_\ell(y,t)  
&:= 
 \int_{\rset^{6N}} \tilde q_\ell(x_t,p_t;y,t) f(x_0,p_0)\Rd x_0\Rd p_0\, ,\\
\tilde q_\ell(x,p;y,t)  &:= 
\sum_n v_\ell^n \eta(y-x^n)
 \big(\frac{M_n|v^n|^2}{2} +\lambda^n(x)\big)  \\
 &\quad +
\sum_{n,m} \int_0^1 \eta\big(y-x^m+s(x^m-x^n)\big) \Rd s(x^n_\ell-x^m_\ell)\times\\
&\quad\times\Big( \frac{p^m}{M_m}\cdot\nabla_{x^m}\lambda^n(x) 
- \sum_j u_j(y,t)\partial_{r^{nm}} \lambda\big(x(r^{12},\ldots, r^{N-1N})\big) \partial_{x_j^n} |x^n-x^m|
\Big)\, ,\\
\end{split}
\end{equation}
and the external energy source as
\begin{equation}\label{Q_def}
P(y,t):= -\int_{\rset^{6N}}\sum_n M_n^{-1}\eta(y-x^n_t)p_t^n
\cdot\nabla_{x^n} V_b(x_t) f(x_0,p_0) \Rd x_0\Rd p_0\, .
\end{equation}

%
%
\section{The conservation laws derived from quantum mechanics}\label{sec_qm}

Irving and Zwanzig \cite{IZ} derived the conservation laws when the particle system
is modeled by the Schr\"odinger equation for a wave function 
$ \Phi:\rset^{3(N+J)}\times [0,\infty)\rightarrow\mathbb C$ satisfying
\begin{equation}\label{schrod_scalar}
\IU \,\partial_t \Phi(\tilde x,t) = \hat \HZ  \Phi(\tilde x,t)\, ,
\end{equation}
with the Hamiltonian
\[
\hat \HZ = -\sum_{n=1}^{N+J} \frac{1}{2\tilde M_n} \Delta_{\tilde x^n} + \nu(\tilde x)\, ,
\]
 based on the nuclei and electron coordinates now
 written together as $\tilde x=(x,\bar x)\in\rset^{3(N+J)}$ with particle masses of nuclei and electrons denoted by $\tilde M_n=M_n, \, n=1,\ldots N$, and $\tilde M_n=1, \, n=N+1,\ldots N+J$, as defined in \eqref{v_def}.
The wave function $\Phi$ is in an appropriate subset of $L^2(\rset^{3(N+J)})$ taking anti-symmetry 
of electron coordinates into account.
Irving and Zwanzig used the Wigner function to establish correspondence between classical and quantum observables.

We use instead the related Weyl quantization, which associates to a (Weyl) symbol $A:\rset^{6(N+J)}\rightarrow \mathbb C$, i.e., a classical observable $A(x,\bar x,p, \bar p)$, an operator on $L^2(\rset^{3(J+N)})$, in fact the Weyl quantization represents an isomorphism between $L^2(\rset^{3(J+N)})$ and the space of Hilbert-Schmidt operators on $L^2(\rset^{3(J+N)})$. 
The Weyl quantized operator $\hat A$ associated with the symbol $A$ in the Schwartz space acting on a function
$\phi\in L^2(\rset^{3(J+N)})$ 
is defined by 
\begin{equation}\label{weyl_def_1M}
\hat A\phi(\tilde x)=\left(\frac{1}{2\pi}\right)^{3(N+J)} \int_{\rset^{6(N+J)}} e^{\IU(\tilde x-\tilde x')\cdot \tilde p}A(\frac{1}{2}(\tilde x+\tilde x'),\tilde p) \phi(\tilde x')
\Rd \tilde x'\Rd \tilde p\, ,
\end{equation}
and the definition is extended to more general symbols $A$ by the standard density arguments. 
For instance, quantization of the  Hamiltonian symbol
\[\HZ(\tilde x,\tilde p)= \sum_{n=1}^{N+J} \frac{ |\tilde p_n|^2}{2\tilde M_n} + \nu(\tilde x)\]
yields the operator $\hat \HZ $.

We recall that the dependence on spin variables is only parametric and it does not enter the quantization procedure since we consider observables and Hamiltonians that do not depend on spin operators, for example modelling systems in absence of magnetic fields, or spin orbit coupling interactions etc. Therefore we omit spin variables $\sigma$, $\bar \sigma$ in the notation. However, we note that the scalar product of wave functions $\phi$, $\psi\in \LSP$  is $\langle \phi,\psi\rangle = \sum_{\sigma}\sum_{\bar\sigma}\int_{\LSP} \phi^*(x,\sigma,\bar x, \bar \sigma) 
\psi(x,\sigma,\bar x, \bar \sigma)\,\Rd x\,\Rd\bar x$.

The Schr\"odinger equation implies the evolution of the wave function $\Phi(\cdot,t)=e^{-\IU t\hat \HZ }\Phi(\cdot,0)$ and consequently 
an observable at time $t$ defined by $\langle \Phi(\cdot, t), \hat A\Phi(\cdot, t)\rangle$ satisfies
\[
\langle \Phi(\cdot, t), \hat A\Phi(\cdot, t)\rangle= 
\langle \Phi(\cdot, 0), e^{\IU t\hat \HZ } \hat Ae^{-\IU t\hat \HZ }\Phi(\cdot, 0)\rangle\PERIOD
\]
By defining the evolution of observables as
\begin{equation}\label{qm_time}
\hat A_t:=e^{\IU t\hat \HZ } \hat Ae^{-\IU t\hat \HZ }
\end{equation}
differentiation implies the Heisenberg-von Neumann equation
\begin{equation*}\label{heissen}
\partial_t \hat A_t = \IU [\hat \HZ ,\hat A_t]\, ,
\end{equation*}
where $[\hat B,\hat C]= \hat B\hat C-\hat C\hat B$ is the commutator.
We also obtain
\begin{equation}\label{A_t_0}
\partial_t \hat A_t = \IU  e^{\IU t\hat \HZ } [\hat \HZ ,\hat A] e^{-\IU t\hat \HZ }
\end{equation}
and $\hat{A}_0=\hat A$.
Let $\hat f$ be the  Weyl quantization of any initial classical density distribution  $f:\rset^{6(N+J)}\to\rset$.
Section \ref{sec:initial} presents a precise
definition of a density symbol related to the given initial data of the macroscopic density, momentum and energy.
The Irving and Zwanzig quantum density observable is then defined by the 
$\LSP$ trace
\begin{equation}\label{qm_dens_def}
\begin{split}
\rho(y,t)&={\TR}(\hat \rho_t \hat f)\\
&:=\sum_{j=1}^{\infty}\langle \Phi_j, \hat \rho_t \hat f\Phi_j\rangle
\end{split}
\end{equation}
where $\{\Phi_j\}_{j=1}^{\infty}$ is a basis of the subspace of $L^2(\rset^{3(N+J)}\times\Sigma\times\bar\Sigma)$ based on the symmetry conditions of fermions and bosons 
and the density observable is the quantization of the density symbol
\[
\begin{split}
\hat \rho_0 &= \big(\sum_{n=1}^{N+J} \tilde M_n\eta(y-\tilde x^n)\big)^{\widehat{ }}\, .\\
\end{split}
\]
The quantum momentum and energy observables are analogously defined as
\begin{equation}\label{p_0_def}
\begin{split}
\hat p_0 &=  \big(\sum_{n=1}^{N+J} \eta(y-\tilde x^n)\tilde p^n\big)^{\widehat{}}\, ,\\
\hat E_0 &= \Big(\sum_{n=1}^{N+J} \eta(y-\tilde x^n)\big(\frac{|\tilde
  p^n|^2}{2\tilde M_n} + 
\nu^n(\tilde x)\big)\Big)^{\widehat{}}\, ,\\
\end{split}
\end{equation}
where $\nu=\sum_{n=1}^{N+J}\nu^n$ is a partition with the potential energy related to each particle, analogous to \eqref{V_n}, now defined as
\[
\nu^n(x,\bar x) =\left\{
\begin{array}{cc}
\frac{1}{2}\sum_{m\ne n} \frac{Z_n Z_m}{|x^n-x^m|}
-\sum_{k=1}^J \frac{ Z_n}{|x^n-\bar x^k|}& \mbox{ for $n\le N$}\\
\frac{1}{2}\sum_{\ell\ne n} 
\frac{1}{|\bar x^n-\bar x^\ell|} & \mbox{ for $n\ge N+1$}\PERIOD\\
\end{array}\right.
\]
Irving and Zwanzig observed the following crucial  property.
\begin{lemma}\label{class_reduct}
For any twice differentiable
scalar symbol that is a polynomial of degree two in the momentum variable
\begin{equation}\label{eq:degtwo}
A(\tilde x,\tilde p):=a_0(\tilde x)+ \sum_{n=1}^{N+J} a_n(\tilde x)\cdot \tilde p^n
+\sum_{m,n=1}^{N+J} a_{n,m}(\tilde x)\tilde p^n\cdot \tilde p^m\in \mathbb{C}
\end{equation}
there holds
\[
\begin{split}
\IU [\hat \HZ ,\hat{A}]&=\big( \nabla_{\tilde p} \HZ (\tilde x,\tilde p) \cdot \nabla_{\tilde x}A(\tilde x,\tilde p)
 - \nabla_{\tilde x} \HZ (\tilde x,\tilde p) \cdot \nabla_{\tilde p}  A(\tilde x,\tilde p)\big)^{\widehat{}}\\
&=: \{\HZ (\tilde x,\tilde p), A(\tilde x,\tilde p)\}^{\widehat{ }}\, ,
\end{split}
\]
where $a_0(\tilde x)\in\rset, a_n(\tilde x)\in\rset^3, a_{n,m}(\tilde x)\in \rset^{{3}\times 3}$.
\end{lemma}
The result is also known in the mathematics literature, cf. \cite[Remarks  2.6.9 and 2.7.6]{martinez}, and we include a proof of the lemma in Section \ref{sec_lemma}, since it is important for this work.
The lemma shows that for a symbol that is a polynomial of degree at most two in the momentum variable
the quantum evolution of the observable reduces to the classical evolution based on  the Poisson bracket.
The quantum observables for density, momentum and energy are  based precisely on symbols
which are degree zero, one  and two, respectively, in the momentum variable.  Irving and Zwanzig therefore
conclude that
the quantum observables satisfy analogous conservation laws as those for classical particle dynamics, namely:
differentiation of the density and using \eqref{A_t_0}, Lemma \ref{class_reduct},  and the definition of $\hat p$ imply the conservation of mass
\begin{equation}\label{commutator_bracket}
\begin{split}
\partial_t\rho(y,t) &= {\TR}\big(\partial_t \hat\rho_t \hat f\big)\\
&=  {\TR}\big(e^{\IU t\hat \HZ }\IU [\hat \HZ ,\hat \rho_0] e^{-\IU t\hat \HZ }\hat f\big)\\
&=  {\TR}\big(e^{\IU t\hat \HZ }\{ \HZ ,\rho_0\}^{\widehat{}} e^{-\IU t\hat \HZ }\hat f\big)\\
&=  -{\TR}\big(e^{\IU t\hat \HZ }{\rm div}_y\hat p_0 e^{-\IU t\hat \HZ }\hat f\big)\\
& =-{\rm div}\Big({\TR}\big(\hat p_t\hat f\big)\Big)\, ,\\
\end{split}
\end{equation}
where the first equality follows by the definition of the macroscopic density \eqref{qm_dens_def}, the second by the Heisenberg-von Neumann dynamics of quantum observables \eqref{A_t_0}, the third by Lemma \ref{class_reduct}, the forth by $\{ \HZ ,\rho_0\}=- \sum_n\nabla\eta(y-x^n)p^n=-{\rm div}_y p_0$ using \eqref{p_0_def} and the fifth by the definition of quantum time evolution \eqref{qm_time}.

Similarly differentiation of the momentum and energy establish the conservation laws
for the momentum 
\begin{equation}\label{zwanzig_momentum}
\begin{split}
\partial_t{\TR}\big(\hat p_t\hat f\big)&
=-\sum_{\ell=1}^3 \partial_{y_\ell} {\TR}\bigg( \Big(
\sum_{n=1}^N \tilde M_n^{-1}\eta(y-\tilde x^n) \tilde p^n \tilde p^n_\ell \\
&\quad 
+\sum_n \sum_m  \int_0^1 \eta\big(y-s\tilde x^n-(1-s)\tilde x^m\big) (\tilde x_\ell^n-\tilde x_\ell^m)\Rd s\\
&\qquad \times\partial_{r^{nm}} \nu\big(\tilde x(r^{12},\ldots, r^{N+J-1N+J})\big) 
\nabla_{\tilde x^n} |\tilde x^n-\tilde x^m|\Big)_t^{\widehat{}}\hat f\bigg)\, \\
\end{split}
\end{equation}
and 
the energy
\begin{equation}\label{zwanzig_energy}
\begin{split}
&\partial_t{\TR}\big(\hat E_t\hat f\big)=
-\sum_{\ell=1}^3 \partial_{y_\ell} {\TR}\bigg(\Big( 
 \sum_n \tilde M^{-1}_n\tilde p_\ell^n \eta(y-\tilde x^n)
 \big(\frac{|\tilde p^n|^2}{2\tilde M_n} + \nu^n(\tilde x)\big) \\
 &\quad +\sum_{n,m} \int_0^1 \eta\big(y-\tilde x^m+s(\tilde x^m-\tilde x^n)\big) \Rd s(\tilde x^n_\ell-\tilde x^m_\ell)
 (\tilde M_m^{-1}\tilde p^m)\cdot\nabla_{\tilde x^m}\nu^n(\tilde x)\Big)_t^{\widehat{}} 
\hat f\bigg)\, .\\
\end{split}
\end{equation}

\subsection{Proof of Lemma \ref{class_reduct}}\label{sec_lemma}
We have the composition rule $\hat B\hat C=\hat D$
where 
\begin{equation}\label{eq:composition}
D = e^{\frac{\IU }{2} (\nabla_{\tilde x'}\cdot\nabla_{\tilde p}-\nabla_{\tilde x}\cdot \nabla_{\tilde p'})}B(\tilde x,\tilde p)
C(\tilde x',\tilde p')\Big|_{
{\tiny \begin{array}{c}
\tilde x=\tilde x'\\
\tilde p=\tilde p'
\end{array}
}} =:B\# C\, ,
\end{equation}
see \cite[Theorem 4.11]{zworski}. Therefore
\[
[\hat \HZ ,\hat{A}]=(\HZ \# A)^{\widehat{}} - ( A\# \HZ )^{\widehat{}}
\]
and, letting $\tilde M$ denote the diagonal matrix with $\tilde M_n$ in the diagonal, we obtain
\[
\begin{split}
&(\nabla_{\tilde x}\cdot\nabla_{\tilde p'}-\nabla_{\tilde x'}\cdot \nabla_{\tilde p})
\big(\HZ (\tilde x,\tilde p) A(\tilde x',\tilde p') - A(\tilde x,\tilde p)\HZ (\tilde x',\tilde p')\big)\\
&= \nabla\nu(\tilde x)\cdot \nabla_{\tilde p} A(\tilde x',\tilde p') 
- \tilde M^{-1}\tilde p\cdot \nabla_{\tilde x}  A(\tilde x',\tilde p')\\
&\quad - \nabla_{\tilde x}  A(\tilde x,\tilde p)\cdot \tilde M^{-1}\tilde p' 
+ \nabla\nu(\tilde x')\cdot \nabla_{\tilde p} A(\tilde x,\tilde p)\\
&= \nabla\nu(\tilde x)\cdot \nabla_{\tilde p}
 A(\tilde x',\tilde p') + \nabla\nu(\tilde x')\cdot \nabla_{\tilde p} A(\tilde x,\tilde p)\\
&\quad - \tilde M^{-1}\tilde p\cdot \nabla_{\tilde x}  A(\tilde x',\tilde p') 
- \tilde M^{-1}\tilde p'\cdot \nabla_{\tilde x} A(\tilde x,\tilde p)=:
\FTERM (\tilde x,\tilde x',\tilde p,\tilde p')\, .
\end{split}
\]
Evaluation at the point $(\tilde x',\tilde p')=(\tilde x,\tilde p)$ yields
\[
\frac{\IU ^2}{2}(\nabla_{\tilde x'}\cdot\nabla_{\tilde p}-\nabla_{\tilde x}\cdot \nabla_{\tilde p'})
\big(\HZ (\tilde x,\tilde p) A(\tilde x',\tilde p') - A(\tilde x,\tilde p)\HZ (\tilde x',\tilde p')
\big)\Big|_{
{\tiny \begin{array}{c}
\tilde x=\tilde x'\\
\tilde p=\tilde p'
\end{array}
}}
=\{ \HZ ,  A\}\, .
\]

The differentiation to the second order becomes
\[
\begin{split}
(\nabla_{\tilde x}\cdot\nabla_{\tilde p'}-\nabla_{\tilde x'}\cdot \nabla_{\tilde p})I(\tilde x,\tilde x',\tilde p,\tilde p')
&= \sum_{m,n}\partial_{\tilde x^m}\partial_{\tilde x^n}\nu(\tilde x) 
\partial_{\tilde p^m}\partial_{\tilde p^n} A(\tilde x',\tilde p')\\
&\quad - \sum_{m,n}\partial_{\tilde x^m}\partial_{\tilde x^n}\nu(\tilde x') 
\partial_{\tilde p^m}\partial_{\tilde p^n} A(\tilde x,\tilde p)\\
&\quad - \sum_n \tilde M_n^{-1} \partial_{\tilde x^n}^2  A(\tilde x,\tilde p)\\
&\quad + \sum_n \tilde M_n^{-1} \partial_{\tilde x^n}^2  A(\tilde x',\tilde p'):=\SECTERM (\tilde x,\tilde x',\tilde p,\tilde p')\, ,\\
\end{split}
\]
so that $\SECTERM (\tilde x,\tilde x,\tilde p,\tilde p)=0$. 

Since the symbol $A$ is a polynomial of degree two in the momentum variable we have
\[
(\nabla_{\tilde x}\cdot\nabla_{p'}-\nabla_{\tilde x'}\cdot \nabla_{\tilde p})\SECTERM (\tilde x,\tilde x',\tilde p,\tilde p')=0
\]
which together with the Taylor expansion of the exponential proves the lemma.

\subsection{Regularization of the observables}
The use of the semiclassical analysis in the next section
requires conditions on the observables,  which are not satisfied for $\eta(y-x^n)$. 
Therefore we replace in the proof of Theorem \ref{thm:md_limit} the function $\eta(y-x^n)$ by
\begin{equation}\label{delta_p_def}
\eta_{\delta_p}(y,x^n,p^n):=\eta(y-x^n)\zeta(|p^n|^2) 
\end{equation}
where $\zeta:\rset\to [0,\infty)$ is a smooth cut-off function satisfying
\begin{equation*}\label{delta_p_def*}
\zeta(q)=\left\{\begin{array}{cc}
1 & \mbox{ for }|q|<1/\delta_p,\\
0 & \mbox{ for }|q|> 2/\delta_p,
\end{array}\right.
\end{equation*}
for a small positive constant $\delta_p>0$\, .

\section{The classical limit of the quantum conservation laws}\label{sec_qm_class}
This section first extends the formulation of quantum conservation laws to the case with matrix valued potentials
in Subsection \ref{sec:ConsMatr},
and then in   Subsection \ref{sec:MDlimit}  this formulation is used to derive molecular dynamics limits of the stress tensor and the heat flux consistent with the quantum conservation laws as formulated in Theorem~\ref{thm:md_limit}.
Therefore we will redefine the notation for the Hamiltonian $H$
and rescale the Weyl quantization 
$A\mapsto \hat A$ and the composition operator $\MP$.

\subsection{Quantum conservation laws with matrix valued potentials}\label{sec:ConsMatr}
The aim here is to consider the quantum evolution where the electron part is considered matrix valued
and the Weyl quantization is  only in the nuclei part. 
For simplicity all nuclei masses are assumed to be equal, $M_n=M$,
and we assume that the nuclei electron mass ratio $M\gg 1$ is large.
The case with individual masses is treated by a change of variables in Section \ref{mass_general}. 
To obtain the classical molecular dynamics limit, as $M\to\infty$, we restrict the electron operator to the finite dimensional $d\times d$ matrix $V$ in \eqref{eigen_v}.
We also change the time scale so that the nuclei dynamics has a limit, when $M\to\infty$, as follows:
the wave function related  to \eqref{schrod_scalar} can be written
as $\Phi:\rset^{3N}\times[0,\infty)\rightarrow \mathbb C^d$  
and it solves
the Schr\"odinger equation
\[
\frac{\IU }{M^{1/2}}\partial_{\tau} \Phi(x,\tau)=\hat H\Phi(x,\tau)\, ,
\]
with the change of the time scale $\tau=M^{-1/2}t$.  
The classical limit obtained as $M\rightarrow\infty$ is well behaved in this time scale, in the sense that
the nuclei move a distance of order one in time one, as we shall see in Theorem \ref{gibbs_corr_thm}. 
The Hamiltonian is now defined as
\[
\hat H=-\frac{1}{2M} \Delta_x {\rm I}+ V(x)+V_b(x){\rm I}
\]
with the Hermitian $d\times d$ matrix $V$ defined in \eqref{V_definitionen}, the external potential $V_b$ satisfying \eqref{vbdry} and ${\rm I}$ the $d\times d$ identity matrix.
In this time scale
the corresponding Heisenberg-von Neumann equation takes the form
\begin{equation*}\label{heisenberg_M}
\partial_{\tau} \hat A_{\tau} = \IU M^{1/2}[\hat H,\hat A_{\tau}]
\end{equation*}
for the $d\times d$ matrix valued symbol $A_{\tau}(x,p)$.
This form of matrix
valued symbols are suitable for studying the $M\rightarrow\infty$ limit
 of the observables we have in the conservation laws, since the corresponding Weyl quantizations are based on
highly oscillatory Fourier integral operators that only use the nuclei coordinates.
We also assume that the eigenvalues 
$\lambda_1(x),\lambda_2(x),\ldots, \lambda_d(x)$ 
of $V(x)$,
defined by \eqref{eigen_v}, satisfy \eqref{A1}-\eqref{A22}.
 The new Weyl quantization takes the form
 \[
 \hat A\phi(x)= \left(\frac{M^{1/2}}{2\pi}\right)^{3N}\int_{\rset^{3N}} \int_{\rset^{3N}}
 e^{\IU M^{1/2}(x-y)\cdot p}A(\XYHALF,p) \phi(y)\,\Rd p\,\Rd y\, ,
 \]
 which differs from \eqref{weyl_def_1M} by the scaling $M^{1/2}$.
 Although this form of matrix valued symbols $H$ and $A$ is useful to obtain
 the classical limit as $M\to\infty$, matrix valued symbols introduce a complication: the important property in Lemma~\ref{class_reduct} that
 the commutator with respect to the conservation observables reduces to the quantization of the Poisson bracket does not hold for matrix valued symbols, since these matrices do not commute in general, unless the symbols are diagonal. 
 A main tool to determine the classical limit is therefore to diagonalize
 $H$ and the observables $A_{\tau}$ in the conservation laws, 
  based on the composition operator $\#$, as follows.

The symbol $C$ for the product of two Weyl operators $\hat A\hat B=\hat C$
is determined by
\begin{equation}\label{eq:composition_2}
C(z,p) = e^{\frac{{\rm i}}{2M^{1/2}} (\nabla_{ z'}\cdot\nabla_{p}-\nabla_{ z}\cdot \nabla_{ p'})}A( z, p)
B( z', p')\Big|_{
{\tiny \begin{array}{c}
 z= z'\\
 p= p'
\end{array}
}} =:(A\# B)(z,p)\, ,
\end{equation}
see \cite[Theorem 4.11]{zworski} which now includes the scaling $M^{1/2}$ as compared to \eqref{eq:composition}. 
 Let $\Psi:\rset^{3N}\rightarrow \mathbb C^{d^2}$ satisfy that  $\Psi(x)$ is a unitary matrix for every $x$ with the Hermitian transpose
 $\Psi^*(x)$, and define
$ {\bar A}:\rset^{3N}\times [0,\infty)\rightarrow \mathbb C^{d^2}$ by
\[
\hat A_{\tau} = \Psi(x) \hat{\bar A}_{\tau}\Psi^*(x)
\]
so that
\[
\hat{\bar A}_{\tau} = \Psi^*(x) \hat{ A}_{\tau}\Psi(x)\, .
\]
Then 
\[
[\hat H,\hat A_{\tau}]= \Psi[\Psi^*\hat H\Psi, \hat{\bar A}_{\tau}]\Psi^*
\]
and consequently
\[
\partial_{\tau}\hat{\bar A}_{\tau}  = \IU M^{1/2}[\Psi^*\hat H\Psi, \hat{\bar A}_{\tau}]\, .
\]
The composition rule \eqref{eq:composition_2} yields $\Psi^*\hat H\Psi=(\Psi^*\#H\#\Psi)^{\widehat{}}$.
The next step is to determine $\Psi$ so that 
\[
\bar{\bar H}:=\Psi^*\#H\#\Psi
\]
is  diagonal or approximate diagonal. Having $\bar{\bar H}$ diagonal implies that $\hat{\bar{\bar H}}$ is diagonal
and then $\hat{\bar A}$ remains diagonal if it is initially diagonal, since then
\[
\begin{split}
\partial_{\tau} \hat{\bar A}_{jk}({\tau})
& = \IU M^{1/2}\big(\hat{\bar{\bar H}}_{jj}\hat{\bar A}_{jk}({\tau}) -
\hat{\bar A}_{jk}({\tau})\hat{\bar{\bar H}}_{kk}\big)=0\, , \quad \mbox{ for } j\ne k.\\
\end{split}
\]

The composition rule \eqref{eq:composition_2} with 
\[
H(x,p)=\big(\frac{|p|^2}{2}+V_b(x)\big) {\rm I} + V(x)
\]
implies that
\[
\begin{split}
\bar{\bar H}&= \Psi^*\# H\# \Psi \\
&= \big(\frac{|p|^2}{2}+V_b(x)\big) {\rm I} + \Psi^*V\Psi + \frac{1}{4M} \nabla\Psi^*\cdot \nabla\Psi\\
&=\Psi^*\Big(\big(\frac{|p|^2}{2}+V_b(x)\big) {\rm I} + V + \frac{1}{4M} \Psi\nabla\Psi^*\cdot \nabla\Psi\Psi^*\Big)\Psi\, ,
\end{split}
\]
as verified in [\cite{KLSS}, Lemma 3.1].
Therefore the aim is to choose the unitary matrix $\Psi$  so that it is an approximate solution to the nonlinear eigenvalue problem
\begin{equation}\label{eigen_nonlin}
\big(V + \frac{1}{4M} \Psi\nabla\Psi^*\cdot \nabla\Psi\Psi^*\big)\Psi= \Psi\bar{\bar\Lambda}
\end{equation}
where $\bar{\bar\Lambda}$ is diagonal.
A solution, $\Psi$, to this nonlinear eigenvalue problem  is an $\mathcal O(M^{-1})$ perturbation of the eigenvectors to $V(x)$
provided the eigenvalues do not cross and $M$ is sufficiently large. The work \cite[(3.18)]{KLSS} shows that \eqref{eigen_nonlin} has an approximate solution $\Psi$ that satisfies
\begin{equation}\label{eigen_nonlin2}
\big(V + \frac{1}{4M} \Psi\nabla\Psi^*\cdot \nabla\Psi\Psi^*\big)\Psi= \Psi\bar\Lambda +\mathcal O(M^{-2})\COMMA
\end{equation}
with $\bar\Lambda$ diagonal, based on the following iteration. Let $\Psi_0=[\psi_1 \ \psi_2\ \ldots\ \psi_d]$ be the matrix of eigenvectors to $V$. The approximate eigenvectors $\Psi$ are the normalized eigenvectors of
\[V+ \frac{1}{4M} \Psi_0\nabla\Psi_0^*\cdot\nabla\Psi_0\Psi_0^*\]
and $\bar\Lambda$ are the corresponding eigenvalues. Since regular perturbation theory shows $\|\Psi-\Psi_0\|_{C^1(\rset^{3N})}=\mathcal O(M^{-1})$ 
we obtain \[\bar{\bar H}(x,p)=\bar H(x,p)+r_0(x)\] where the remainder $d\times d$ matrix $r_0(x)$ is small 
\begin{equation}\label{r_0_est}
\|r_0\|_{L^\infty(\rset^{N})}=\mathcal O(M^{-2})\COMMA
\end{equation}
with the diagonal matrix
\begin{equation}\label{bar_H}
\bar H(x,p):= \Big(\frac{|p|^2}{2}+V_b(x)\Big) {\rm I} +\bar\Lambda(x)\, .
\end{equation}

We also need a partition of the eigenvalues $\bar\Lambda=\sum_{n=1}^N\bar\Lambda^n$ related to the potential energy for each particle similar to \eqref{lambda_n_def}, now including also the small nonlinear part. In fact, also the nonlinear part
has a natural composition into particle contributions, now based on the sensitivity of the eigenvectors with respect to position $x^n$.  With $V^n$ defined by \eqref{V_definitionen} we have
\begin{equation*}
V + \frac{1}{4M} \Psi\nabla\Psi^*\cdot \nabla\Psi\Psi^*
=\sum_{n=1}^N \big(V^n + \frac{1}{4M} \Psi\nabla_{x^n}\Psi^*\cdot \nabla_{x^n}\Psi\Psi^*\big)
\end{equation*}
and define for $n=1,\ldots, N$
\begin{equation}\label{bar_lambda_n}
\begin{split}
\bar\lambda_k^n &:= \langle \Psi_k, (V^n + \frac{1}{4M} \Psi\nabla_{x^n}\Psi^*\cdot \nabla_{x^n}\Psi\Psi^*)\Psi_k\rangle\, , k=1,\ldots, d\, ,\\
\bar\Lambda^n &= {\rm diag}(\bar\lambda^n_1, \ldots, \bar\lambda_d^n)\, ,
\end{split}
\end{equation}
which implies
\[
\begin{split}
\bar\lambda_k&= \sum_{n=1}^N \bar\lambda_k^n\, ,\\
\bar\Lambda &= \sum_{n=1}^N \bar\Lambda_k^n\, .
\end{split}
\]

We will use the observables defining density, momentum and energy as follows.
Let
\[
\bar\rho_0(x,y)=\sum_{n=1}^N\eta(y-x^n){\rm I}
\]
and as before its time evolution is determined by the Heisenberg-von Neumann equation
\[
\partial_{\tau} \hat{\bar\rho}_{\tau}= iM^{1/2}[\hat{\bar{\bar H}},\hat{\bar \rho}_{\tau}]\, ,
\]
with the solution
\[
\hat{\bar\rho}_{\tau} = e^{\IU {\tau}M^{1/2}\hat{\bar{\bar H}}}\hat{\bar\rho}_0 e^{-\IU {\tau}M^{1/2}\hat{\bar{\bar H}}}\, ,
\]
which shows that the time evolution also can be written as
\begin{equation}\label{rho_ekv_1}
\partial_{\tau} \hat{\bar\rho}_{\tau}= \IU M^{1/2}e^{\IU {\tau}M^{1/2}\hat{\bar{\bar H}}}
[\hat{\bar{\bar H}},\hat{\bar \rho}_0]e^{-\IU {\tau}M^{1/2}\hat{\bar{\bar H}}}\, .
\end{equation}

The momentum and energy density symbols are defined as
\[
\begin{split}
\bar p_0 &= \sum_{n=1}^N\eta(y-x^n)p^n{\rm I}\, ,\\
\bar E_0 &= \sum_{n=1}^N\eta(y-x^n)\big( \frac{|p^n|^2}{2}{\rm I} + \bar\Lambda^n(x)\big)\, .
\end{split}
\]

The next step is to derive the quantum conservation/balance laws by studying the evalution of the observables for density, momentum and energy.
Let $\hat f={\Psi}\hat{\bar f}{\Psi}^*$ be the Weyl quantization of a given density symbol $\bar f(x,p)\in\mathbb R^{d\times d}$
as described precisely in Section \ref{sec:initial}.
\begin{lemma}\label{lemma_mass}
Assume that the eigenvalues $\lambda_k, \ k=1,\ldots, d$ of $V$ are distinct and there are positive constants $C$ and $c$ such that
\[
\begin{split}
\sum_{|\alpha|\le 2}\|\partial^\alpha \psi_k\|_{L^2(\rset^{3N})}
+\sum_{|\alpha|\le 2}\|\partial^\alpha \lambda_k\|_{L^2(\rset^{3N})}+
\|\bar f\|_{L^2(\rset^{6N})}+\|\bar\rho_0\|_{L^2(\rset^{6N})} &\le C\COMMA\\
\|\bar f\|_{L^1(\rset^{6N})} &> c\COMMA\\
\end{split}
\]
then
\begin{equation}\label{qm_mass_cons_1}
\begin{split}
\partial_{\tau} {\TR}(\Psi\hat{\bar\rho}_{\tau}\Psi^* \hat f)
&=\partial_{\tau} {\TR}(\hat{\bar\rho}_{\tau} \hat{\bar f} )\COMMA\\
\partial_{\tau} {\TR}(\hat{\bar\rho}_{\tau} \hat{\bar f} ) &=-{\rm div}\big({\TR}(\hat{\bar p}_{\tau} \hat{\bar f})\big)  +\mathcal O(M^{-3/2}) \TR(\hat{\bar f})\COMMA
\end{split}
\end{equation}
as $M\to\infty$.
\end{lemma}
As compared to \eqref{commutator_bracket},
the quantum continuity equation  \eqref{qm_mass_cons_1} includes a $\mathcal O(M^{-3/2})$ remainder term, due to non perfect diagonalization in \eqref{eigen_nonlin2}.

\begin{proof}
We have
\[
[\hat{\bar{\bar H}},\hat{\bar\rho}_0]=
[\hat{\bar{ H}},\hat{\bar\rho}_0] + [\hat r_0,\hat{\bar\rho}_0],
\]
where $\|r_0\|_{L^{\infty}(\rset^{3N})}=\mathcal O(M^{-2})$ by \eqref{r_0_est}.

The diagonal form of $\bar H$ combined with the property that the symbols
$\bar\rho_0$, $\bar p_0 $, $\bar E_0$ and $\bar H$ are polynomials of degree at most two as functions of the momentum coordinate imply by Lemma \ref{class_reduct}, rewritten in the new scaling, 
the reduction of the corresponding quantum commutators to classical Poisson brackets. 
In the new scaling Lemma \ref{class_reduct}
takes the form: assume that $\bar H$ and $\bar A$
are diagonal matrices where each component of $\bar A$ is a polynomial of degree at most two in $p$, as in \eqref{eq:degtwo}, then
${\rm i}M^{1/2}[\hat{\bar H},\hat{\bar A}]=\{\bar H,\bar A\}^{\widehat{}}$\, . 
Therefore, we have as in \eqref{commutator_bracket} 
\[
\begin{split}
\IU M^{1/2} [\hat{\bar H}, \hat{\bar\rho}_0]& =
\{\bar H,\bar\rho_0\}^{\widehat{}}
=-{\rm div}_y \hat{\bar p}_0\, ,\\ 
\end{split}
\]
and by \eqref{rho_ekv_1}
\begin{equation}\label{cons_law1}
\begin{split}
 \partial_{\tau} \hat{\bar\rho}_{\tau} &= - {\rm div}_y \hat{\bar p}_{\tau}+
 \IU M^{1/2}e^{\IU {\tau}M^{1/2}\hat{\bar{\bar H}}}[\hat{r}_0,\hat{\bar \rho}_0]e^{-\IU {\tau}M^{1/2}\hat{\bar{\bar H}}}\, .\\
 \end{split}
\end{equation}
By taking the trace in \eqref{cons_law1}
the conservation law for the mass becomes
\begin{equation}\label{qm_mass}
\begin{split}
\partial_{\tau} {\TR}(\Psi\hat{\bar\rho}_{\tau}\Psi^* \hat f)
&= 
\partial_{\tau} {\TR}(\hat{\bar\rho}_{\tau}\Psi^* \hat f\Psi)\\
&=\partial_{\tau} {\TR}(\hat{\bar\rho}_{\tau} \hat{\bar f} )\\
&=
-{\rm div}\big({\TR}(\hat{\bar p}_{\tau} \hat{\bar f})\big)
+\TR\big(\IU M^{1/2}e^{\IU {\tau}M^{1/2}\hat{\bar{\bar H}}}[\hat{r}_0,\hat{\bar \rho}_0]e^{-\IU {\tau}M^{1/2}\hat{\bar{\bar H}}}\hat{\bar f}\big)\
\, , 
\end{split}
\end{equation}

The next step is to estimate the remainder term including $r_0$.
Cauchy's inequality in the Hilbert-Schmidt inner product $\TR(\hat A^* \hat B)$ implies that the remainder term has the estimate
\begin{equation}\label{remainder11}
\begin{split}
&|\TR\big(\IU M^{1/2}e^{\IU {\tau}M^{1/2}\hat{\bar{\bar H}}}[\hat{r}_0,\hat{\bar \rho}_0]e^{-\IU {\tau}M^{1/2}\hat{\bar{\bar H}}}\hat{\bar f}\big)|\\
&\le M^{1/2}\Big(\TR\big([\hat{r}_0,\hat{\bar \rho}_0]^2\big)\ 
\TR\big((e^{-\IU {\tau}M^{1/2}\hat{\bar{\bar H}}}\hat{\bar f}e^{\IU {\tau}M^{1/2}\hat{\bar{\bar H}}})^*e^{-\IU {\tau}M^{1/2}\hat{\bar{\bar H}}}\hat{\bar f}e^{\IU {\tau}M^{1/2}\hat{\bar{\bar H}}}\big)\Big)^{1/2}\\
&=M^{1/2}\big(\TR([\hat{r}_0,\hat{\bar \rho}_0]^2)
\TR\big(\hat{\bar f}^*\hat{\bar f}\big)\big)^{1/2}\\
&=M^{1/2}\Big(\TR\big(({r_0}\#{\bar \rho}_0-\bar\rho_0\# r_0)^{\widehat{}}
\ )^2\big)
\TR\big(\hat{\bar f}^*\hat{\bar f}\big)\Big)^{1/2}\PERIOD\\
\end{split}
\end{equation}
The Weyl quantization satisfies
\[
\begin{split}
\TR(\OPER{A}) &=\NORMFAC^{3N}\int_{\mathbb R^{6N}}  \TR\big(A(z)\big) \Rd z\COMMA\\
\TR(\OPER{A}\OPER{B})&=\NORMFAC^{3N}\int_{\mathbb R^{6N}}  \TR\big(A(z)B(z)\big) \Rd z\COMMA
\end{split}
\]
where $A(z)B(z)$ is the matrix product of the two $d\times d$ matrices $A(z)$ and $B(z)$, with the second trace acting on matrices, see \cite{teufel} and \cite[Lemma 3.1]{KLSS}. This isometry between Hilbert-Schmidt operators and $L^2(\rset^{6N},\mathbb C^{d\times d})$ functions also extends the Weyl quantization from symbols in the Schwartz class to $L^2(\rset^{6N},\mathbb C^{d\times d})$, see \cite{teufel}.
Lemma \ref{comp_lemma} implies
\[
\begin{split}
\TR\Big(\big(({r_0}\#{\bar \rho}_0-\bar\rho_0\# r_0)^{\widehat{}}
\ \big)^2\Big)&=\NORMFAC^{3N}\int_{6N}\TR\big(({r_0}\#{\bar \rho}_0-\bar\rho_0\# r_0)^2\big) {\rm d} z\\
&\le 4\NORMFAC^{3N}\|r_0\|^2_{L^\infty(\rset^{3N})}\|\bar\rho_0\|^2_{L^2(\rset^{6N})}
\end{split}
\]
and we obtain by \eqref{r_0_est} and \eqref{remainder11}
\[
|\TR\big(\IU M^{1/2}e^{\IU {\tau}M^{1/2}\hat{\bar{\bar H}}}[\hat{r}_0,\hat{\bar \rho}_0]e^{-\IU {\tau}M^{1/2}\hat{\bar{\bar H}}}\hat{\bar f}\big)|=\mathcal O(M^{-3/2})\TR(\hat{\bar f})\COMMA
\]
which by \eqref{qm_mass} and \eqref{remainder11} proves the lemma.
\end{proof}

The work \cite[Lemma 3.11]{KLSS} proves
\begin{lemma}\label{comp_lemma}
Assume  
$D:\rset^{6N}\rightarrow \cset^{d\times d}$
belong to $L^2(\rset^{6N})$
and $A:\rset^{3N}\rightarrow \cset^{d\times d}$ depends only on the $x$-coordinate (or only on the $p$-coordinate) and is 
bounded in $L^\infty(\rset^{3N})$ then
\begin{equation}\label{CD}
\begin{split}
\|A\MP D\|_{L^2(\rset^{6N})}&\le \|A\|_{L^\infty(\rset^{3N})}\|D\|_{L^2(\rset^{3N})}\COMMA\\
\|D\MP  A\|_{L^2(\rset^{6N})}&\le \|A\|_{L^\infty(\rset^{3N})}\|D\|_{L^2(\rset^{3N})}\PERIOD\\
\end{split}
\end{equation}
\end{lemma}

The conservation of momentum and energy are also based on
the reduction from commutators to Poisson brackets, in Lemma \ref{class_reduct}, as follows
\begin{equation}\label{cons_law2}
\begin{split}
\partial_{\tau} \hat{\bar p}_{\tau} &= -{\rm div} \big(\sum_n \eta(y-x^n) p^n\otimes p^n{\rm I}\big)^{\widehat{}}_{\tau}\\
&\quad - {\rm div}\Big(\sum_n \sum_k  \int_0^1 \eta\big(y-sx^n-(1-s)x^k\big) (x_\ell^n-x_\ell^k)\Rd s\\
&\qquad \times\partial_{r^{nk}} \bar\Lambda^n\big(x(r^{12},\ldots, r^{N-1N})\big) 
\nabla_{x^n} |x^n-x^k| \Big)^{\widehat{}}_{\tau} \\
&\quad - \Big(\underbrace{\sum_n \eta(y-x^n) \nabla_{x^n} V_b(x)}_{=:-\bar F(x,y)}\Big)^{ \widehat{}}_{\tau} 
 +\underbrace{\IU M^{1/2}e^{\IU {\tau}M^{1/2}\hat{\bar{\bar H}}}[\hat{r}_0,\hat{\bar p}_0]e^{-\IU {\tau}M^{1/2}\hat{\bar{\bar H}}}}_{=:R_1}\\
&:=-{\rm div}\hat{\bar S}_{\tau} +\hat{\bar F}_\tau + R_1\, ,\\
\partial_{\tau} \hat{\bar E}_{\tau} &= -{\rm div} 
\Big(\sum_n \eta(y-x^n)p^n \big( \frac{|p^n|^2}{2}{\rm I} + \bar\Lambda^n(x)\big)\Big)^{\widehat{}}_{\tau}\\
&\quad - {\rm div} \Big(\sum_n \int_0^1 \eta\big(y-x^m+ s(x^m-x^n)\big)\Rd s
(x^n-x^m) p^m\cdot \nabla_{x^m} \bar\Lambda^n) \Big)^{\widehat{}}_{\tau}\\
&\quad -\Big(\underbrace{\sum_n M_n^{-1}\eta(y-x^n)p^n
\cdot\nabla_{x^n} V_b(x)}_{=:-\bar P(x,y)}\Big)^{\widehat{}}_{\tau}
+ \underbrace{\IU M^{1/2}e^{\IU {\tau}M^{1/2}\hat{\bar{\bar H}}}[\hat{r}_0,\hat{\bar E}_0]e^{-\IU {\tau}M^{1/2}\hat{\bar{\bar H}}}}_{=:R_2}\\
&=: -{\rm div}\hat{\bar Q}_{\tau} +\hat{\bar P}_\tau+ R_2\, .\\
\end{split}
\end{equation}
As for the conservation of mass in Lemma \ref{lemma_mass} we obtain the conservation/balance law for the momentum
\begin{equation}\label{qm_momentum}
\partial_{\tau} {\TR}(\Psi\hat{\bar p}_{\tau}\Psi^* \hat f)
= -{\rm div}\big({\TR}(\Psi\hat{\bar S}_{\tau}\Psi^* \hat f)\big)
+{\TR}(\Psi\hat{\bar F}_{\tau}\Psi^* \hat f)
+\mathcal O(M^{-3/2})\TR(\hat{\bar f})\, , 
\end{equation}
and the conservation/balance law for the energy
\begin{equation}\label{qm_energy}
\partial_{\tau} {\TR}(\hat\Psi\hat{\bar E}_{\tau}\hat\Psi^* \hat f)
= -{\rm div}\big({\TR}(\hat\Psi\hat{\bar Q}_{\tau}\hat\Psi^* \hat f)\big)
+{\TR}(\hat\Psi\hat{\bar P}_{\tau}\hat\Psi^* \hat f)
+\mathcal O(M^{-3/2})\TR(\hat{\bar f})\, , 
\end{equation}
provided $\bar p_0,\bar S_0,\bar F_0,\bar E_0,\bar Q_0,\bar P_0$ are all bounded in $L^2(\rset^{6N})$.

In Section~\ref{sec:initial} we will motivate
an initial density $\hat f=\hat\Psi \hat{\bar f}\hat \Psi^*$ as a local grand canonical Gibbs density, where $\bar f$ is diagonal, with the local temperature and chemical potential determined by the macroscopic density and energy. That is, $\hat f$ is diagonalized by the same transformation as $\hat H$.
The traces in the quantum conservation laws \eqref{qm_mass_cons_1} \eqref{qm_momentum} and \eqref{qm_energy} can then be written as 
\[
\begin{split}
{\TR}(\Psi \hat{\bar A}_{\tau}\Psi^* \hat f)&= 
{\TR}( \hat{ A}_{\tau} \hat f)
= {\TR}( \hat{\bar A}_{\tau}\Psi^* \hat f\Psi)
={\TR}( \hat{\bar A}_{\tau}\hat{\bar f})
\end{split}
\]
where $\bar A$ is diagonal and equal to $\bar\rho, \bar p, \bar E, \bar S$ and $\bar Q$, respectively. The next section presents the classical limit of these traces.


\subsection{The molecular dynamics limit of the quantum conservation laws}\label{sec:MDlimit}
 The work \cite{KLSS}
 proves  in Theorem 3.7 the following which provides the classical limit
of the quantum observables in the conservation laws. A related result, with different assumptions, is in \cite{teufel}. The proof is
 based on Weyl's law, see \cite{teufel,KLSS} namely that quantum observables have the classical representation
 \begin{equation}\label{Wlaw}
 \frac{{\TR}( \hat{\bar A}_{0}\hat{\bar f})}{{\TR}( \hat{\bar f})}=
 \frac{\sum_{j=1}^d\int_{\rset^{6N}}\bar A_{jj}(0,z)\bar f_{jj}(z)\Rd z}{\sum_{j=1}^d\int_{\rset^{6N}}\bar f_{jj}(z)\Rd z}
 \end{equation}
 for any $\bar A_{jj}(0,\cdot)\in L^2(\rset^{6N})$ and $\bar f_{jj}\in L^2(\rset^{6N})\cap L^1(\rset^{6N})$.

\begin{theorem}\label{gibbs_corr_thm}
 Assume that $V$ 
 satisfies the coercivity  condition \eqref{A2}, the $d\times d$ matrices $\bA_0$ and $\bar f$ are diagonal, the $d\times d$ matrix-valued Hamiltonian $H$ has distinct eigenvalues, and that there is a constant $C$ such that
\begin{equation*}\label{R77}
\begin{split}
\sum_{|\alpha|\le 2}\|\partial^\alpha_x \psi_k\|_{L^\infty(\rset^{3N})} &\le C\COMMA\quad k=1,\ldots,d\COMMA\\
\max_i\sum_{|\alpha|\le 3}\|\partial^\alpha_x \partial_{x_i}\lambda_j\|_{L^\infty(\rset^{3N})} &\le C\COMMA\\
\sum_{|\alpha|\le 3}
\|\partial_z^\alpha\bA_{jj}(0,\cdot)\|_{L^2(\rset^{6N})}  &\le C\COMMA\\
\| \bar f 
\|_{L^2(\rset^{6N})} 
&\le C\COMMA\\
\end{split}
\end{equation*} 
hold, then there is a constant $c$, depending on $C$, such that the canonical ensemble average satisfies the error estimate
\begin{equation*}\label{G_corr_unif1}
\begin{split}
&\Big|\frac{\TR\big(\OPER{\bar A}_\tau  \widehat{\bar f} 
\big)}{
\TR(\OPERW{ \bar f 
})}
 - \sum_{j=1}^d \int_{\rset^{6N}} 
\frac{\bA_{jj}(0,z^j_\tau (z_0))
\bar f_{jj}(z_0)
}{\sum_{k=1}^d\int_{\rset^{6N}}
\bar f_{kk}(z)
\Rd z} \Rd z_0\Big|
\le  cM^{-1}\COMMA
\end{split}
\end{equation*}
as the mass ratio $M\rightarrow\infty$, 
where $z^j_{\tau}=(x_{\tau},p_{\tau})$ is the solution to the Hamiltonian system
\begin{equation}\label{HS}
\begin{split}
\dot x_{\tau} &= p_{\tau}\\
\dot p_{\tau} &= -\nabla\bar \lambda_j(x_{\tau}) -\nabla V_b(x_\tau), \quad {\tau}>0\, ,
\end{split}
\end{equation}
based on the Hamiltonian $\bar H_{jj}(z)=|p|^2/2 + \bar\lambda_j(x)+ V_b(x)$,
with initial data $(x_0,p_0)=z_0$.
\end{theorem}

We note that the classical limit can be written 
\begin{equation*}
\sum_{j=1}^d \int_{\rset^{6N}} 
\frac{\bA_{jj}\big(0,z^j_\tau (z_0)\big)
\bar f_{jj}(z_0)
}{\sum_{k=1}^d\int_{\rset^{6N}}
\bar f_{kk}(z)
\Rd z} \Rd z_0
=  \sum_{j=1}^d \int_{\rset^{6N}} q_j^*\bar A_{jj}\big(0,z^j_\tau (z_0)\big)  
\frac{\bar f_{jj}(z_0)}{\int_{\rset^{6N}}\bar f_{jj}(z)\Rd z} \Rd z_0
\end{equation*}
where the probability, $q_j^*$, to be in electron state $j$ is
\begin{equation}\label{q_def}
q_j^*:=\frac{\int_{\mathbb R^{6N}} \bar f_{jj}(z)
\Rd z}{
\sum_{k=1}^d\int_{\mathbb R^{6N}}
\bar f_{kk}(z')
\Rd z'}\, ,\quad j=1,\ldots, d\, .
\end{equation}

To apply Theorem \ref{gibbs_corr_thm} to the quantum observables
\eqref{qm_mass_cons_1}, \eqref{qm_momentum} and \eqref{qm_energy} for macroscopic density, momentum and energy the momentum variable
$\eta$ needs to be regularized, since e.g. the momentum symbol $\bar p_0=\sum_n\eta(y-x^n)p^n{\rm I}$ is not in $L^2(\rset^{6N})$. Therefore we regularize all symbols by replacing $\eta(y-x^n)$ by  $\eta_{\delta_p}:=\eta(y-x^n)\zeta(|p^n|^2)$, given in \eqref{delta_p_def}, and denote 
the quantum observables using $\eta_{\delta_p}$ instead of $\eta$ in \eqref{qm_mass_cons_1}, \eqref{qm_momentum} and \eqref{qm_energy}
as $\TR(\hat{\bar\rho}_\tau\hat{\bar f})_{\delta_p}$ (replacing $\TR(\hat{\bar\rho}_\tau\hat{\bar f})$ and similarly for the other observables).
\begin{assumption}\label{assumption_delta_p}
Assume that, for any regularization $\delta_a$ and  dimension $d$, the quantum observables
\begin{equation}\label{tr_delta}
\TR(\hat{\bar\rho}_\tau\hat{\bar f})_{\delta_p}, \TR(\hat{\bar p}_\tau\hat{\bar f})_{\delta_p}, \TR(\hat{\bar S}_\tau\hat{\bar f})_{\delta_p},
\TR(\hat{\bar F}_\tau\hat{\bar f})_{\delta_p}, \TR(\hat{\bar E}_\tau\hat{\bar f})_{\delta_p}, \TR(\hat{\bar Q}_\tau\hat{\bar f})_{\delta_p}, \TR(\hat{\bar P}_\tau\hat{\bar f})_{\delta_p}
\end{equation}
{and their derivatives with respect to $\tau$ and $y$ have limits as 
$\delta_p\to 0+$,   with the limits based on $\eta=\eta_0$.}
\end{assumption}

Theorem \ref{gibbs_corr_thm} and the assumed continuous dependence on the regularization parameters 
can be used to show a consistency result, namely that as the nuclei electron mass ratio $M$ tends to infinity in
the  quantum conservation laws 
\eqref{qm_mass_cons_1}, \eqref{qm_momentum} and \eqref{qm_energy},
using the splitting $p_{\tau}^n=v_{\tau}^n+u(y,\tau)$ for the fluxes $\bar S$ and $\bar Q$,
we obtain the following limit in the form of a macroscopic conservation/balance law based on a certain stress tensor and heat flux defined by molecular dynamics including several electron eigenvalues.  
We note that the diagonal terms in the
flux terms $\bar S$ and $\bar Q$
are the same as in the classical dynamics \eqref{mom_class} and \eqref{en_class}.
\begin{theorem}\label{thm:md_limit}
Assume that the approximate electron operator $V$ satisfies\eqref{A1}-\eqref{A22}
and Assumption~\ref{assumption_delta_p} 
and the assumptions in Theorem \ref{gibbs_corr_thm} hold, with $\bar A$ diagonal and equal to
$\bar \rho_0, \bar p_0, \bar E_0, \bar S_0$ and  $\bar Q_0$,
then as the nuclei-electron mass ratio $M\to\infty$ the quantum conservation laws \eqref{qm_mass_cons_1}, \eqref{qm_momentum} and \eqref{qm_energy} have a classical molecular dynamics limit that satisfies
\begin{equation}\label{cons_class_r}
\begin{split}
\partial_{\tau}\rho(y,t) + \sum_{\ell=1}^3\partial_{y_\ell}\big(\rho(y,t)u_\ell(y,\tau)\big) &=0\, ,\\
\partial_{\tau}\big(\rho(y,{\tau})u_j(y,{\tau})\big) + \sum_{\ell=1}^3\partial_{y_\ell}\big(\rho(y,{\tau})u_j(y,{\tau})u_\ell(y,{\tau})- \sigma_{\ell j}(y,{\tau})\big) &=F_j(y,t)\, ,\\
\partial_{\tau} E(y,{\tau}) + \sum_{\ell=1}^3\partial_{y_\ell}\big( E(y,{\tau})u_\ell(y,{\tau}) + q_\ell(y,{\tau})- \sum_j\sigma_{\ell j}(y,{\tau})u_j(y,{\tau}) \big) &= P(y,t)\, ,\\
\end{split}
\end{equation}
where
\begin{equation}\label{md_list}
\begin{split}
\rho(y,{\tau})&=\sum_{j=1}^\infty q^*_j\int_{\rset^{6N}} \bar\rho_{jj}(z^j_{\tau},y)\bar f_{jj}(z_0)\Rd z_0\, ,\\
u(y,{\tau}) &= \sum_j q^*_j\int_{\rset^{6N}} \bar p_{jj}(z^j_{\tau},y)\bar f_{jj}(z_0)\Rd z_0/\rho(y,{\tau})\, ,\\
E(y,{\tau}) &=\sum_j q^*_j\int_{\rset^{6N}} \bar E_{jj}(z^j_{\tau},y)\bar f_{jj}(z_0)\Rd z_0\, ,\\
\sigma(y,{\tau}) &=\sum_j q^*_j\int_{\rset^{6N}} \bar\sigma_{}(z^j_{\tau};y,{\tau})\bar f_{jj}(z_0)\Rd z_0\, ,\\
q(y,{\tau}) &=\sum_j q^*_j\int_{\rset^{6N}} \bar q_{}(z^j_{\tau};y,{\tau})\bar f_{jj}(z_0)\Rd z_0\, ,\\
F(y,\tau) &= \sum_j q^*_j\int_{\rset^{6N}} \bar F(z^j_{\tau},y,{\tau})\bar f_{jj}(z_0)\Rd z_0\, ,\\
P(y,\tau) &= \sum_j q^*_j\int_{\rset^{6N}} \bar P(z^j_{\tau},y,{\tau})\bar f_{jj}(z_0)\Rd z_0\, ,\\
q_j^*&=\frac{\int_{\mathbb R^{6N}} \bar f_{jj}(z)
\Rd z}{
\sum_{k=1}^\infty\int_{\mathbb R^{6N}}
\bar f_{kk}(z')
\Rd z'}\, ,
\end{split}
\end{equation}
and  $\bar\sigma(z^j;y,{\tau})$ and $\bar q(z^j;y,{\tau})$ are defined in \eqref{sigma_def} and \eqref{heat_def}, respectively, now using $\lambda=\bar\lambda_j$, and the matrix valued symbols 
\begin{equation}\label{matrix_cons_def}
\begin{split}
\bar{\rho}_0&=\sum_n\eta(y-x^n){\rm I}\, ,\\
\bar{p}_0&=\sum_n\eta(y-x^n)p^n{\rm I}\, ,\\
\bar{E}_0&=\sum_n\eta(y-x^n)\big(\frac{|p^n|^2}{2} {\rm I} + \bar\Lambda^n(x)\big)\, ,\\
\bar F_0 &=\sum_n \eta(y-x^n) \nabla_{x^n} V_b(x){\rm I}\, ,\\
\bar P_0 &=\sum_n M_n^{-1}\eta(y-x^n)p^n
\cdot\nabla_{x^n} V_b(x){\rm I}\, .
\end{split}
\end{equation}
\end{theorem}

\begin{proof}
The quantum observables \eqref{tr_delta}, based on the approximate electron operator $V$ with finite $d$ and positive parameter
$\delta_a$,
satisfy the assumptions in Theorem \ref{gibbs_corr_thm} with $\eta$  regularized as $\eta_{\delta_p}$. 
These regularized quantum observables therefore have a classical limit as $M\to\infty$: given small positive  $(\delta_a,\delta_p)$
and large finite $d,M$, the leading order terms in the classical approximation is by Theorem~\ref{gibbs_corr_thm} arbitrary close to the terms in \eqref{md_list}
while the error term $c/M$ can be made sufficiently small using sufficiently large $M$.

To verify that the observables satisfy the conservation laws, we need to take the limit $\delta_p\to 0+$ in both the quantum observables \eqref{tr_delta} and the leading order classical term \eqref{md_list}, since the quantum conservation laws \eqref{qm_mass_cons_1}, \eqref{qm_momentum} and \eqref{qm_energy} are given with  $\delta_p=0$.
As $\delta_p\to 0+$, the regularized observables in \eqref{tr_delta}, based on the positive $\delta_a$,
converge by Assumption  \ref{assumption_delta_p} to the observables that satisfy the conservation laws \eqref{qm_mass_cons_1}, \eqref{qm_momentum} and \eqref{qm_energy}. Combined with the continuous dependence on $\delta_p\to 0+$ in the leading order terms of the molecular dynamics approximation \eqref{md_list} we obtain the consistency results
\eqref{cons_class_r}-\eqref{matrix_cons_def}.
\end{proof}

The result of Theorem~\ref{thm:md_limit} still depends on the regularization parameters $\delta_a$ and $d$, while the
ab initio model corresponds to $\delta_a=0$ and $d=\infty$. The molecular dynamics observables in Theorem~\ref{thm:md_limit} are consistent with the ab initio quantum model in the sense that if the molecular dynamics observables in Theorem \ref{thm:md_limit} and the $M\to \infty$ limits of the quantum observables in \eqref{tr_delta} have limits as $\delta_a\to 0+$ and $d\to\infty$, with the limits based on $\delta_a=0$ and $d=\infty$, then the corresponding molecular dynamics and quantum limits are equal.

Theorem \ref{thm:md_limit} proves in particular that the observables are determined
by a weighted average with the probability $q^*_j$ to be in state $j$. Using this probability model,
the conservation laws \eqref{cons_class_r} are consistent with the derivation of the conservation laws from the classical dynamics in Section \ref{sec_classical}.


\subsection{General nuclei masses}\label{mass_general}
The general case
of individual nuclei masses and a diagonal mass matrix $M$
can be treated by rescaling the nuclei position coordinates as
$M_1^{1/2}x'=M^{1/2}x$, which transforms the Hamiltonian into
\[-(2M_1)^{-1}{\rm I}\Delta_{x'}  +  V(M_1^{1/2}M^{-1/2}x')+ V_b(M_1^{1/2}M^{-1/2}x')\, .\]
In these transformed coordinates the classical limit is by Theorem \ref{gibbs_corr_thm} based on
Lemma \ref{class_reduct} applied to $\{\bar H(x',p'),\bar A\big(x(x'),p(p')\big)\}$ with $\bar A$ given by
the conservation variables \eqref{matrix_cons_def}, which take the form
 $\{\bar H_{jj}(x',p'),\bar A_{jj}\big(x(x'),p(p')\big)\}
=\{|p|^2/2+ M_1M^{-1} \bar\lambda_j(x), \bar A_{jj}(x,p)\}$ in the original variables.
Therefore the transformed Hamiltonian system 
\begin{equation*}\label{HS1}
\begin{split}
\dot x'_{\tau} &= p'_{\tau}\\
\dot p'_{\tau} &= -\nabla_{x'}\bar \lambda_j\big(x_{\tau}(x')\big) -\nabla V_b\big(x_\tau(x')\big), \quad {\tau}>0\, ,
\end{split}
\end{equation*}
provides
the untransformed system
\begin{equation*}\label{HS2}
\begin{split}
\dot x_{\tau} &= p_{\tau}\\
\dot p_{\tau} &= -M_1M^{-1}\big(\nabla_x\bar \lambda_j(x_{\tau}) +\nabla V_b(x_\tau)\big), \quad {\tau}>0\, ,
\end{split}
\end{equation*}
which is equivalent to the standard form
\begin{equation*}\label{HS3}
\begin{split}
\dot x_{\tau} &= M_1M^{-1}q_{\tau}\\
\dot q_{\tau} &= -\nabla_x\bar \lambda_j(x_{\tau}) -\nabla V_b(x_\tau), \quad {\tau}>0\, ,
\end{split}
\end{equation*}
with individual masses in the diagonal mass matrix $M$.

\subsection{Coinciding eigenvalues}\label{sec_conical}
The assumption on distinct eigenvalues in Theorem \ref{gibbs_corr_thm}
is used in the proof to obtain differentiable eigenvalues and eigenvectors by regular perturbation theory in \eqref{eigen_nonlin2}.
%
Here we indicate how to relax the assumption on distinct eigenvalues for a special example. 

The first step is to perturb $\tVd$ to obtain distinct eigenvalues. We have $V=\Psi\Lambda\Psi^*$
where $\Psi$ is the matrix with the eigenvectors as columns and $\Lambda$ is the diagonal matrix of eigenvectors $(\lambda_1,\ldots, \lambda_d)$.
Assume, for example, there are points $x'$ such that eigenvalues $\tLm_1(x')=\tLm_2(x')$
coincide and the other pairs of eigenvalues do not coincide. 
Since we have $\lambda_1\le \lambda_2<\lambda_3$ we can regularize 
$\breve\lambda_i(x) :=\int_{\rset^{3N}}\lambda_i(y)\frac{e^{-|x-y|^2/2\delta_e}}{(2\pi \delta_e)^{d/2}}{\rm d}y$ for $i=1,2$ so that for small $\delta_e>0$ we obtain distinct eigenvalues
$\breve\lambda_1< \breve\lambda_2<\lambda_3$, which are  smooth
since $\lambda_i$ is Lipschitz continuous.
Replace $V$ by the new matrix $\breve V:=\Psi\breve\Lambda\Psi^*$
where $\breve\Lambda$ is the diagonal matrix with $\lambda_i$ replaced by $\breve\lambda_i$ for $i=1,2$. Since the eigenvalues of $\breve V$
are distinct and smooth.

It remains
to study  the eigenvectors.
The eigenvectors corresponding to $\lambda_1$ and $\lambda_2$ are typically not continuous at conical intersections $x$, where $\lambda_1(x)=\lambda_2(x)$, see e.g. \cite[Section 12.2.3]{tannor}. The proof of Theorem \ref{gibbs_corr_thm} uses the regularity of the eigenvectors of $V$ only 
to obtain \eqref{eigen_nonlin2}.
The approximate nonlinear diagonalization becomes surprisingly simple near a conical intersection for the case of a real valued $2\times 2$ matrix $V=\left[\begin{array}{cc}v_{11} & v_{12}\\v_{12} &v_{22}\end{array}\right]$. Its eigenvalues are $\lambda_{\pm}= \frac{v_{11}+v_{22}}{2} \pm r$ where $r:=\sqrt{(\frac{v_{11}-v_{22}}{2})^2+v_{12}^2}$ with the corresponding eigenvectors 
\[\Psi=\left[\begin{array}{cc}\cos\alpha/2 & \sin\alpha/2\\-\sin\alpha/2 &\cos\alpha/2\end{array}\right]\PERIOD\]
Here $r(\cos\alpha,\sin\alpha)=(\frac{v_{11}-v_{22}}{2},v_{12})$, so
\[
V=\frac{v_{11}+v_{22}}{2}\, {\rm I} 
+r\left[\begin{array}{cc}\cos\alpha & \sin\alpha\\\sin\alpha &-\cos\alpha\end{array}\right]\COMMA
\]
and when $r(x)=0$ there is a  conical intersection with $\lambda_+(x)=\lambda_-(x)$.
Let $\tilde v:=\frac{v_{11}-v_{22}}{2}$, then $\tan\alpha=v_{12}/\tilde v$ so $\alpha=\tan^{-1}v_{12}/\tilde v+ n\pi$ and $\nabla\alpha(x)=\frac{\tilde v\nabla v_{12}-v_{12}\nabla\tilde v}{\tilde v^2 + v_{12}^2}$, which implies
\[
\nabla\Psi^*\cdot\nabla\Psi=\frac{|\nabla\alpha|^2}{4}
\left[\begin{array}{cc}-\sin\alpha/2 & \cos\alpha/2\\-\cos\alpha/2 &-\sin\alpha/2\end{array}\right]^*
\left[\begin{array}{cc}-\sin\alpha/2 & \cos\alpha/2\\-\cos\alpha/2 &-\sin\alpha/2\end{array}\right]
=\frac{|\nabla\alpha|^2}{4}{\rm I}
\]
and consequently 
\[
\big(V + \frac{1}{4M} \Psi\nabla\Psi^*\cdot \nabla\Psi\Psi^*\big)\Psi= \Psi(\Lambda + \frac{|\nabla\alpha|^2}{16M}{\rm I})=\Psi\bar\Lambda
\]
in fact solves the nonlinear eigenvalue problem \eqref{eigen_nonlin} exactly.
We note that the nonlinear eigenvalue 
$\bar\lambda_k$ has the additional term $|\nabla\alpha|^2/(16M)$ which is
large if $r$ is small, i.e. in the potential landscape of $\bar\lambda_k$ there is a mountain around the conical intersection that becomes lower as M gets larger. 
These observations could be the first step to extend Theorem \ref{gibbs_corr_thm} to include conical intersections.

\section{The initial particle density}\label{sec:initial}
Assume we know the initial data 
$\big(\rho(\cdot,0),\rho u(\cdot,0),E(\cdot,0)\big)$
for the macroscopic conservation laws, although in practise this data can be hard to determine, e.g. for the flow  in a river. Current molecular dynamics simulations can only use a small fraction of the number of particles in a real system. Therefore we need an initial particle density that is related to a larger ensemble. We seek a density that has the property that the marginal distribution of a  subsystem 
weakly coupled to a larger heat bath system is the same as the whole system. Under certain assumptions stated in \cite{KLSS} the classical Gibbs density is the only density with this property.  Given the local values of the macroscopic conserved variables the goal here is therefore to determine a local grand canonical Gibbs density
\begin{equation}\label{dgibbs}
\bar f_{jj}(x,p;y)\sim e^{-\mathcal H(x,p,j;y)/T(y)}
\end{equation}
where
\[
\mathcal H(x,p,j;y):=\sum_{n=1}^N\eta(y-x^n)\big(\underbrace{\bar H^n(x,p,j)}_{=\frac{|p^n|^2}{2M_n}+\bar\lambda^n_j(x)}- M_n\mu(y)\big)\, ,
\]
based on local values of the temperature $T(y)$ and the chemical potential $\mu(y)$, such that
\begin{equation}\label{initial_repr}
\left(\begin{array}{c}
\rho(y,0)\\ \rho u(y,0)\\ E(y,0)
\end{array}
\right) =\big({\TR}(\hat{\bar  f})\big)^{-1}{\TR} 
\left(\left(\begin{array}{c} \hat{\bar \rho}_0\\  \hat{\bar p}_0\\ \hat{\bar E}_0
\end{array}
\right) 
 \hat{\bar f}\right)\, .
\end{equation}

Weyl's law  given by the
quantum-classical representation \eqref{Wlaw} combined with \eqref{initial_repr} show that the equation 
\[
\rho u(y,0)=\frac{\int_{\rset^{6N}}
\sum_{n=1}^N \eta(y-x^n) p^n
\sum_{j=1}^d  \bar f_{jj}(x,p)\Rd x\Rd p}{
\int_{\rset^{6N}}\sum_{j=1}^d  \bar f_{jj}(x,p)\Rd x\Rd p
}
\]
defines the initial velocity $u(y,0)$. It remains to  verify if varying $T(y)$ and $\mu(y)$ yield large enough sets to match the initial data for $\rho$ and $E$. 

If $|\mu(y)|\gg 1$ we roughly get $\mathcal H\simeq -\sum_{n=1}^N \eta(y-x^n)\mu(y)$. Laplace principle implies that
as $\mu(y)\to -\infty$ the grand canonical density will sample the minimum of the observable 
$\sum_{n=1}^N \eta(y-x^n)$
and as $\mu(y)\to\infty$ the microscopic particle density will sample the maximum
of $\sum_{n=1}^N \eta(y-x^n)$. Therefore varying $\mu(y)$ from $-\infty$ to $\infty$  will change the local density  
\[\rho(y,0)=\frac{\int_{\rset^{6N}}\sum_{n=1}^N \eta(y-x^n) M_n
\sum_{j=1}^d  \bar f_{jj}(x,p)\Rd x \Rd p}{
\int_{\rset^{6N}}\sum_{j=1}^d  \bar f_{jj}(x,p)\Rd x\Rd p
}
\]
from nearly vacuum to arbitrary high macroscopic density.

The temperature is related to the microscopic kinetic energy
and we have
\[
E(y,0)= \frac{\int_{\rset^{6N}}\sum_{n=1}^N \sum_{j=1}^d\eta(y-x^n) 
\big( \frac{|p^n|^2}{2M_n} + \bar\lambda_j^n(x)\big)
  \bar f_{jj}(x,p)\Rd x \Rd p}{
\int_{\rset^{6N}}  \bar f_{jj}(x,p)\Rd x\Rd p
}\, .
\]
We see that the factor 
$\eta(y-x^n) 
 \frac{|p^n|^2}{2M_n}e^{-\eta(y-x^n)|p^n|^2/(T(y)2M_n)}$
upon integration with respect to $p^n$ will be proportional to $T(y)$. The other terms in $E(y,0)$ will have upper and lower bounds uniform in $T$. Therefore by varying the temperature a large open set of macroscopic energies can be attained.

Finally, we note that minimizing the entropy $\sum_{j=1}^d\int_{\rset^{6N}}\bar f_{jj}(z) \log \bar f_{jj}(z) \Rd z$ under the constraints that the value of the macroscopic density  is $\rho(y,0)$ and the macroscopic energy is $E(y,0)$
yields the probability density 
\[
\bar f_{jj}(z)=ce^{-\sum_{n=1}^N \eta(y-x^n)M_n\mu_0(y) -\sum_{n=1}^N \eta(y-x^n)\bar H^n(z,j)\mu_1(y)}
=c e^{-\mathcal H(z,j;y)/T(y)}
\]
for $c=1/\sum_{j=1}^d\int_{\rset^{6N}}e^{-\mathcal H(z,j;y)/T(y)}\Rd z$ with the Lagrange multipliers $\mu_1(y)=1/T(y)$ and $\mu_0(y)=\mu(y)/T(y)$. The Gibbs density \eqref{dgibbs} is therefore consistent with this constrained minimization.

\section{The partial derivatives $\partial_{r^{jk}}\tilde\lambda(\tilde x)$}\label{sec:derivatives}
The function $\tilde\lambda$ is defined by \eqref{eq:tildelambda} on
the set $M\subset\rset^{N(N-1)/2}$ consisting of all vectors
$r:=(r^{12},r^{13},\ldots,r^{N-1 N})$ such that there exist particle positions
$x^i\in\rset^3$, $i=1,\ldots,N$ and $r^{ij}=\lvert x^i-x^j\rvert$ for
$1\leq i<j\leq N$. 
The goal here is to define the gradient
$\nabla_r\tilde\lambda$ such that the chain rule
$\nabla\lambda(x)=\big(\partial r/\partial
x\big)^T\nabla_r\tilde\lambda(r)$ is valid on $M$. We therefore need
to solve the linear equation $Av=b$ for $v=\nabla_r\tilde\lambda$,
where we have used the notation $A=\big(\partial r/\partial
x\big)^T$ and $b=\nabla\lambda$. Since this is an underdetermined
linear system we choose the solution $v$ that minimizes the $\ell^2$
norm. The stationary point to the Lagrangian $L(v,y)=\lvert v
\rvert^2/2 + y\cdot(Av-b)$, where $y$ is the Lagrange multiplier, has
the solution 
\begin{equation}\label{eq:leastsquares}
v=A^T(AA^T)^{-1} b.
\end{equation}
Since generically the matrix $A$
has full rank, equation \eqref{eq:leastsquares} contains a computable
expression for $v=\nabla_r\tilde\lambda$, which can be used in the
expression for the stress tensor in \eqref{sigma_def}.

\end{document}